# Growth and Properties of Superconducting MgB$_2$ Thin Films

Michio Naito and Kenji Ueda

NTT Basic Research Laboratories, NTT Corporation,

3-1 Wakamiya, Morinosato, Atsugi, Kanagawa 243-0198, Japan

**Abstract**

This review article describes the developments over the last 30 months in the thin film growth and junction fabrication of superconducting MgB$_2$, including a brief summary the chemistry and physics of MgB$_2$. The most serious problem is high Mg vapor pressure required for the phase stability. This problem makes *in-situ* film growth difficult. So, most of the initial efforts were performed by two-step growth, namely *ex-situ* post-annealing. Later, *in-situ* physical vapor deposition and recently *in-situ* chemical vapor deposition were reported. Each growth method is described in detail and compared. The past efforts at fabricating MgB$_2$ junctions are summarized, and the future prospects for MgB$_2$ superconducting electronics are briefly described.

**Contents**

# 1. Introduction

Magnesium diboride (MgB$_2$) is a binary intermetallic compound with a simple crystal structure (so called AlB$_2$ structure). MgB$_2$ is not a new compound, its synthesis was reported in the mid 1950's, and it is even commercially available. However, the superconductivity of MgB$_2$ was overlooked for many years until it was announced in January, 2001 by Akimitsu's group [1]. This remarkably simple compound has a $T_c$ of around 40 K.

MgB$_2$ has had considerable impact on the superconductivity field for a number of reasons. The first is that the $T_c$ of 40 K is the highest superconducting transition temperature of intermetallic compounds (the second highest is 23 K in Nb$_3$Ge [2] and YPd$_2$B$_2$C [3]). The second reason is that MgB$_2$ seems to be a phonon-mediated BCS superconductor, which was demonstrated by the isotope effect reported immediately after the first announcement of superconductivity [4, 5]. The $T_c$ of 40 K is actually higher than the BCS limit (~30 K). For high-$T_c$ cuprates, whose $T_c$ is considerably higher than the BCS limit, it has been claimed that the superconducting mechanism should be exotic and involve spin-mediated pairing rather than phonon-mediated pairing. Therefore, the phonon-mediated superconductivity with $T_c$ ~ 40 K in MgB$_2$ requires old theories to be reexamined. The third reason is that MgB$_2$ has many properties that make it very attractive for superconducting applications, especially electronics applications. Unlike high-$T_c$ cuprates [6], MgB$_2$ may be suitable, despite its lower $T_c$, for fabricating good Josephson junctions. This is because it has less anisotropy, fewer material complexities, fewer interface problems, and a longer coherence length ($\xi$ ~5 nm). One possible demerit of MgB$_2$ for practical applications may be that there is no material variety. Any substitution leads to a $T_c$ reduction, and there are no superconducting materials with $T_c > 30$ K in the neighborhood.   This is in contrast to high-$T_c$ cuprates, where more than 300 superconducting compounds have been discovered.

The scope of this article is to review the developments over the last 30 months in the thin film growth and junction fabrication of $MgB_2$ toward superconducting electronics applications.   This review article is organized as follows.   Before dealing with the main topics, we briefly summarize the chemistry and physics of $MgB_2$ in Sections 2 and 3.   Section 4 provides a general description of the thin film growth of $MgB_2$, focusing on a comparison of the two limiting methods, namely two-step versus *in-situ* growth.   The results of our MBE growth of $MgB_2$ films are described in Section 5.   Section 6 summarizes past efforts on the fabrication of $MgB_2$ junctions, including our own results.   Section 7 presents the summary and describes the future prospects for $MgB_2$ superconducting electronics.

**2. Chemistry of superconducting $MgB_2$**

2.1. Crystal structure

Magnesium diboride, $MgB_2$, has a hexagonal $AlB_2$ structure.   Figure 1 is a schematic representation of this crystal structure [1].   The structure consists of hexagonal-close-packed (hcp) layers of Mg atoms alternating with graphite-like honeycomb layers of B atoms.   The lattice constants are a = 3.086 Å and c = 3.524 Å.

2.2. Phase diagram and thermodynamics

It is important to know the phase diagrams of $MgB_2$ for thin film growth.  This is because there is a very large imbalance between the vapor pressures of Mg and B: Mg is a highly volatile element, in contrast, B is an element with a high melting temperature. Hence one can expect difficulties with regard to the *in-situ* preparation of $MgB_2$ thin films. In contrast, we may expect the volatility of Mg to assist the automatic composition adjustment as a result of the self-limiting adsorption of Mg, as found with the film growth of III-V and II-VI semiconductors.

Liu *et al.* calculated the phase diagrams based on a thermodynamic analysis of the Mg-B binary system [7]. In the Mg-B system, there are three intermediate compounds, $MgB_2$, $MgB_4$, and $MgB_7$, in addition to the gas, liquid, and solid magnesium phases and the solid boron phase. Evaluating the Gibbs free energy of each phase, the phase equilibria are calculated. The resultant phase diagrams are reproduced in Fig. 2. In Fig. 2, the temperature-composition phase diagrams for the Mg-B system are plotted at different Mg pressures ($P_{Mg}$): (a) 1 atm, (b) 1 Torr, and (c) 1 mTorr. The results are summarized as follows.

(1) For an atomic Mg:B ratio, $x_{Mg}/x_B$, greater than 1:2, the $MgB_2$ phase coexists with either solid Mg or liquid Mg or gas Mg below the decomposition temperature ($T_d$). Liquid Mg does not exist when $P_{Mg}$ is lower than the triple-point pressure of Mg (650°C, 2.93 Torr)

(2) $P_{Mg}$ has a significant influence on the decomposition temperature ($T_d$): $T_d$ = 1545°C at 1 atm, 912°C at 1 Torr, and 603°C at 1 mTorr. Above $T_d$, $MgB_2$ decomposes into a mixture of $MgB_4$ and Mg vapor.

Based on thermodynamic considerations, it is preferable to perform the thin film growth of $MgB_2$ within a "window" for $MgB_2$ + Mg-gas, in which $MgB_2$ does not decompose and excess Mg does not condense on the $MgB_2$ surface. Within this "window", MgB$_2$ film growth can be as easy as GaAs film growth. Such a convenient "window" is best illustrated by the pressure-temperature phase diagram shown in Fig. 3.

2.3 Kinetics

In Fig. 3, the MgB$_2$ + Mg-gas window is fairly large, indicating that MgB$_2$ film growth *would* not be difficult. However, this has proved to be untrue. According to Fig. 3, for example, with $P_{Mg}$ = 10$^{-4}$ - 10$^{-6}$ Torr, which is compatible with vacuum deposition techniques, MgB$_2$ would be formed at a growth temperature of ~400°C. In reality, however, MgB$_2$ films do not form under such growth conditions. The thermodynamics only tells us that, once it forms, MgB$_2$ does not decompose in the phase stability field. However, it does not guarantee phase formation in this field since the kinetics has to be taken into consideration in the crystallization. According to our experience, with $P_{Mg}$ = 10$^{-4}$ - 10$^{-6}$ Torr and $T_s$ ~ 400°C, Mg reevaporates from the surface before Mg is trapped by B, resulting in no MgB$_2$ phase formation.

Figure 4 shows the vapor pressure curve of elemental Mg compared with the MgB$_2$ decomposition curve taken from Fig. 3. In this figure, we have also included another vertical scale (on the right) that shows the Mg flux ($F$ [Å/sec]), which is converted from the Mg partial pressure by the following relation,

$$F = P / \sqrt{2\pi m k_B T} \quad \text{or} \quad F \text{ [Å/sec]} = 6.9043 \times 10^5 \times P \text{ [Torr]}$$

where $m$ is the Mg mass and $T$ is taken as ~300°C, which is a typical MBE growth temperature. The vapor pressure curve is significantly higher than the decomposition curve. With an Mg vapor pressure of 10$^{-6}$ Torr corresponding to an Mg flux of 0.69

Å/sec, free Mg is lost above 250°C because the reevaporation rate is higher than the impinging rate. At the same Mg vapor pressure, Mg in the MgB$_2$ lattice is lost above 420°C. Above 250°C, there will be no net accumulation of Mg on the substrate unless Mg reacts with B to form MgB$_2$.

So far, we have discussed the growth kinetics of MgB$_2$. The decomposition kinetics is another important aspect of thin film preparation, especially in the post-deposition annealing process. Fan *et al.* undertook experimental studies on the MgB$_2$ decomposition process by measuring the evaporation rate of Mg from MgB$_2$ in a vacuum using either a quartz crystal microbalance or a residual gas analyzer [8]. Their results are included in Fig. 4, which represents the evaporation pressure (desorption rate) of Mg from MgB$_2$ as a function of MgB$_2$ temperature. This curve is well below the thermodynamic decomposition curve. The evaporation coefficient ($\alpha$), which is defined as the ratio of the observed Mg rate to the thermodynamically predicted rate, is $10^{-4}$. This indicates that the decomposition reaction of MgB$_2$ is very slow because of a large kinetic barrier.

## 3. Physics of superconducting MgB$_2$
3.1 Isotope effect and phonons

In this section, we provide a brief survey of the physical properties of MgB$_2$. We start with the isotope effect and phonons in MgB$_2$. The isotope effect is one of the fundamental tests for the superconducting mechanism. In a conventional BCS superconductor, where Cooper pairing is mediated by phonons, the isotope exponent ($\alpha$ in $T_c \sim M^{-\alpha}$) is 1/2. Immediately after the first announcement of superconductivity in MgB$_2$, the isotope effect for MgB$_2$ was measured by Bud'ko *et al.* and Hinks *et al* [4, 5]. The B isotope exponent $\alpha_B$ ($\alpha_B = -\Delta \ln T_c / \Delta \ln M_B$) in MgB$_2$ is as large as 0.26 to 0.30, whereas the Mg isotope effect is small ($\alpha_{Mg} \sim 0.02$). The results strongly support phonon-mediated superconductivity in MgB$_2$ with B atom vibrations substantially involved. The total isotope exponent, ($\alpha_t = \alpha_B + \alpha_{Mg}$), is ~0.32, which is slightly reduced from the ideal value of 1/2. This can be explained either by strong Coulomb repulsion or by large anharmonicity of B atom vibrations.

The pressure effect of $T_c$ also supports the phonon-mediated superconductivity in MgB$_2$ [9, 10]. As is often the case with conventional superconductors, the $T_c$ decreases with pressure in MgB$_2$ up to the highest pressure studied. This behavior can be understood by assuming that the electron-phonon interaction decreases as the lattice stiffens.

MgB$_2$ appears to be a conventional, BCS phonon-mediated *sp* superconductor. We must next ask why a superconducting transition temperature as high as 40 K can be achieved in this compound. The first intuition is that this compound is made up of light mass elements, resulting in high phonon frequencies ($\omega_{ph} \sim M^{1/2}$) that set the $T_c$ temperature scale in BCS theory. The phonon density of states of MgB$_2$ has been obtained experimentally from the inelastic neutron scattering measurements on polycrystalline samples undertaken by Osborn *et al.* [11] and Yildirim *et al* [12]. The two sets of results are in basic agreement, and also confirm the above speculation. The experimental data indicate that the phonon density of states extends as high as 100 meV as shown in Fig. 5. There are two bands of phonons: one below 40 meV corresponding primarily to acoustic modes and the other above 50 meV corresponding to optic branches mostly involving the boron motions. The high frequency phonons are also confirmed by specific heat measurements, which provided a Debye temperature for MgB$_2$ of over 1000 K. The electron-phonon coupling constant ($\lambda$) can be estimated from these data based on the McMillan or Allen-Dynes equation. Using the Allen-Dynes equation with $\omega_{ln}$ = 57.9 meV, Osborn *et al.* obtained $\lambda$ ~0.9, indicating an intermediate coupling regime. In their calculation, they did not assume that a specific phonon mode makes a significant contribution to the superconductivity. The band calculations [13, 14], however, predicted that the specific boron in-plane vibration mode (bond-stretching mode, see Fig. 5(c)) is strongly coupled to electrons (B $2p\sigma$ band) near the Fermi level. This is because this phonon mode changes the B orbital overlap, thereby causing strong electron-phonon coupling. This mode has $E_{2g}$ symmetry with an energy of ~75 meV. Assuming that this mode dominates electron-phonon coupling in MgB$_2$, the $\lambda$ value can be calculated as 0.9-1.0 by a deformation potential approximation, again indicating an intermediate coupling regime.

3.2 Two-gap superconductivity

The superconducting energy gap ($\Delta$) of MgB$_2$ has been inferred from various spectroscopic probes (tunneling, far-infrared optics, photoemission, point contact) as well as from the temperature dependence of physical quantities (specific heat, microwave surface resistance, penetration depth) at low temperatures. There was a large scattering from 2 to 8 meV for $\Delta$ in early studies. This large scattering was initially attributed to surface effects (degraded surface layers) as in the case of high-$T_c$ cuprates. However, later experiments have pointed towards the existence of multiple gaps. Two-gap (or two-band) superconductivity is not a new concept, and was suggested soon after the development of the BCS theory [15]. In the past, some *s-d* metals were thought to fall into the category of multi-band superconductors, generally in the form of small deviations from BCS predictions. Later Binnig *et al.* claimed two-band superconductivity for Nb-doped SrTiO$_3$ on the basis of tunneling spectroscopy [16].

With MgB$_2$, the clearest evidence for two-gap superconductivity was obtained by specific heat measurements [17, 18]. Specific heat is a bulk property, and is not sensitive to surface conditions. The specific heat ($C$) of MgB$_2$ significantly deviates from the standard BCS behavior as shown in Fig. 6. First, there is a significant excess $C$ in the vicinity of $T_c$/5 (~8 K). Second, an exponential fit of $C(T)$ in the $T \ll T_c$ region indicates a small superconducting gap with $2\Delta/k_B T_c \sim 1.2$. These two observations can be interpreted as indicating the existence of a second small gap. In order to interpret the behavior of $C(T)$ in MgB$_2$, Bouquet *et al.* proposed the following semi-quantitative analysis. In their model, $C(T)$ is the sum of the contributions of two bands of large and small superconducting gaps ($\Delta_L$ and $\Delta_S$) with their relative weights $x$ and 1- $x$ : $C(T) = C_L(T) + C_S(T)$. Then each $C_i(T)$ (i = L or S) can be calculated assuming a BCS temperature dependence for $\Delta_i$. The calculated $C(T)$ agrees well with the experimentally obtained specific heat over the whole $T$ to $T_c$ range with fitted parameters of $2\Delta_l/k_B T_c \sim 4.0$, $2\Delta_s/k_B T_c \sim 1.2$, and $x \sim 0.55$. Manzano *et al.* reached a similar conclusion from the temperature dependence of the London penetration depth ($\lambda(T)$) shown in Fig. 7 [19]. Magnetic penetration depth experiments are not strictly bulk measurements, but they probe the sample to a typical depth of $\lambda$ = 1000-2000 Å. Nevertheless, $\lambda$ is large compared with the typical sampling depths of many spectroscopic experiments such as tunneling, photoemission, etc. A two-gap analysis for $\lambda$(T) gave $2\Delta_l/k_B T_c \sim 3.9-4.5$, $2\Delta_s/k_B T_c \sim 1.5$, and $x \sim 0.55$, which is in good agreement with the above specific heat results.

Some spectroscopic measurements also provided similar results. Tsuda *et al.* studied the superconducting gap of $MgB_2$ using high-resolution photoemission spectroscopy [20]. Their spectrum in the superconducting state showed a complex gap structure in contrast to that expected from a simple isotropic gap. This spectrum can be well reproduced using the weighted sum of two Dynes functions with gap values of 5.6 meV ($2\Delta_l/k_BT_c \sim 3.6$) and 1.7 meV ($2\Delta_s/k_BT_c \sim 1.1$). The spectroscopy of Cu/ $MgB_2$ point-contact junctions reported by Szabo *et al.* gave $\Delta_L = 7$ meV and $\Delta_S = 2.8$ meV [21]. Vacuum tunneling spectroscopy using $MgB_2$ as an STM tip against an atomically flat 2H-$NbSe_2$ undertaken by Giubileo *et al.* gave $\Delta_L = 7.8$ meV and $\Delta_S = 3.8$ meV [22].

The postulation of the two-gap superconductivity of $MgB_2$ can be justified from the band calculations [23, 24]. Band calculations reported by several authors gave essentially identical results, and pointed out the following features.

(1) Mg is substantially ionized, and $MgB_2$ can be well described as the ionic form $Mg^{2+}(B_2)^{2-}$.

(2) The bands near the Fermi level mainly derive from two distinct sets of boron orbitals: $sp^2$ ($\sigma$) states and $p_z$ ($\pi$) states. The $\sigma$ bands are 2D in character and form cylindrical Fermi surfaces, whereas the $\pi$ bands have more of a 3D character owing to a substantial *c*-axis transfer integral and form 3D tubular networks of Fermi surfaces.

(3) The $\sigma$ bands (2D FS) interact strongly with the $E_{2g}$ phonon mode and are predicted to have a large 2D superconducting gap. In contrast, the $\pi$ bands have weaker electron-phonon interaction and their superconducting gap is about 3 times smaller in magnitude.

3.3 Normal-state properties

The basic physical properties of $MgB_2$ were studied initially using polycrystalline samples. The results from these initial studies differed from group to group. This is especially true for transport properties, which are most sensitive to weak links at grain boundaries. Soon after, however, several groups began to measure small single crystals grown using high-pressure furnaces or single-crystalline films using a two-step method [25]. This resulted in making the experimental data more consistent. Most of the physical properties are relatively normal, which is in contrast to those of high-$T_c$ cuprates. For example, the resistivity as shown in Fig. 8 can be well fitted by the Bloch-Grüneisen formula with a very high Debye temperature of 1000 ± 100 K [26]. The resistivity value of $MgB_2$ at room temperature is 5-6 μΩcm, which can be compared to ρ(300K) of ~2 μΩcm for Cu, ~15 μΩcm for Nb, ~80 μΩcm for $Nb_3Sn$, and 100 – 150 μΩcm for YBCO. However, the low-temperature resistivity still differs from group to group: 0.3 μΩcm to 3 μΩcm [26, 27]. The simplest explanation attributes this large variation to the amount of impurity scattering. However, it has also been suspected of being caused by remnant Mg providing an extra conduction path, leading to a resistivity lower than the intrinsic value. Further study is required to confirm whether the intrinsic conduction dominates the observed resistivity.

The Hall coefficient ($R_H$) is positive for magnetic field ($H$) parallel to the $c$-axis, and negative for $H$ perpendicular to $c$ as shown in Fig. 9 [28]. Both $R_H$ values show substantial temperature dependence. This complex $R_H$ behavior indicates the presence of two bands with different anisotropies, which is qualitatively consistent with the band calculations. A rough estimate of the carrier density from $R_H$ ($H//c$) gives 2-3 x $10^{23}$ $cm^{-3}$, which can be compared to 6.7 x $10^{22}$ $cm^{-3}$ obtained from two free electrons per unit cell. This carrier density is similar to that of conventional metals and two orders-of-magnitude larger than that of YBCO.

The magnetic susceptibility is nearly temperature independent (Pauli paramagnetic behavior), except for the Curie tail, which is caused by a small quantity of magnetic impurities [29]. The Pauli paramagnetic susceptibility is ~16.2 x $10^{-6}$ emu/mole, corresponding to $N(E_F)$ = 0.249 states/eV-cell-spin, which is about 30% smaller than the calculated value, $N(E_F)$ = 0.36 states/eV-cell-spin.

3.4 Superconducting properties

Most of the fundamental superconducting properties were summarized in a previous review by Buzea and Yamashita [30]. So we only include a brief description here. The upper critical field of $MgB_2$ is anisotropic because of its layered structure. Recent measurements of $H_{c2}$ on single crystals from several groups are in fairly good agreement [31-33]. One typical set of data is reproduced in Fig. 10 [32]. The $H_{c2}^{//c}$ value for the magnetic field parallel to the $c$ axis is low, about 3 to 4 T, even at $T = 0$ K. The $H_{c2}^{//ab}$ value for $H$ parallel to the $ab$ plane is substantially higher than $H_{c2}^{//c}$, but reaches only 15 to 20 T at $T = 0$ K. The resultant anisotropy ($H_{c2}^{//c} / H_{c2}^{//ab}$) is 4-5. The GL coherence lengths at $T = 0$ K can be calculated from the equations, $H_{c2}^{//c} = \phi_0/(2\pi\xi_{ab}^2)$ and $H_{c2}^{//c} = \phi_0/(2\pi\xi_{ab}\xi_c)$, as $\xi_{ab} = 10\pm2$ nm, $\xi_c = 2$-3 nm. The mean free path ($l_{ab}$) of $MgB_2$ single crystals is as large as 50-100 nm [34], which indicates that $MgB_2$ is in the clean limit ($l_{ab}/\xi_{ab} \gg 1$). The rather low value of $H_{c2}$ is unfavorable for the practical application of $MgB_2$. However, the artificial introduction of impurities or defects may lead to a reduction in $\xi$, and thereby an increase in $H_{c2}$ by the conventional GLAG impurity effect. In fact, it seems to be a general trend that high-resistivity samples of $MgB_2$ seem to have substantially higher $H_{c2}$ than the value quoted above for low-resistivity single crystals.

The lower critical field $H_{c1}$ is also anisotropic. There are much fewer reports on anisotropic $H_{c1}$ than on $H_{c2}$. One report from Xu *et al.* provided that $H_{c1}^{//c}(0) = 27.2$ mT and $H_{c1}^{//ab}(0) = 38.4$ mT [35]. Although the average $H_{c1}$ value is in rough agreement with that obtained from polycrystalline samples (Li *et al.*) [36], this anisotropy is the opposite of that expected from the conventional anisotropic GL equation: $H_{c1}^{//c}$ should be larger than $H_{c1}^{//ab}$. From the polycrystalline data for $H_{c1}$, the magnetic penetration depth can be roughly estimated as $\lambda(0) \sim 170$ nm using $H_{c1} = \phi_0 \ln \kappa / (2\pi\lambda^2)$. For comparison, $\lambda_{ab} = 110\text{-}130$ nm and $\lambda_c = 210\text{-}280$ nm were obtained from a frequency shift in a resonant LC circuit for single crystal $MgB_2$ specimens. For single crystalline films prepared by two-step growth, $\lambda_{ab} = 150$ nm was obtained by a mutual inductance method [37], and $\lambda_{ab} = 100$ nm by a sapphire dielectric resonator technique [38].

## 4. General description of $MgB_2$ film growth

### 4.1 Two-step versus in-situ growth

In this section, we provide a general description of $MgB_2$ film growth. There are two complicated problems related to the preparation of superconducting $MgB_2$ films: the high sensitivity of Mg to oxidation and the high vapor pressure of Mg required for phase stability. The former problem can be avoided by depositing $MgB_2$ films in an ultra high vacuum or in a reducing atmosphere containing hydrogen. The latter problem is more serious, and there are two ways to overcome it. One is to prepare the film under a high Mg vapor pressure in a confined container at a high temperature, and the other is to prepare the film at a low temperature. The first approach is employed in the "two-step growth" technique, in which the first step is the deposition of amorphous B (or Mg-B composite) precursors, and the second step is the annealing of the precursors at an elevated temperature with Mg vapor usually in an evacuated Nb, Ta or quartz tube. The second approach is employed with the "*in-situ* growth" technique. Each approach has its own merits and demerits. Two-step growth produces good crystalline films with superconducting properties comparable to those of high-quality sintered specimens although it cannot be used to fabricate Josephson junctions or multilayers. In contrast, *in-situ* growth can produce only poor crystalline films that have a slightly lower $T_c$ (typically 35 K) than the bulk value, but this approach makes multilayer deposition feasible.

4. 2 Two-step growth

The two-step growth process is as follows [39-43]. First, amorphous B (or Mg-B composite) precursors are deposited on substrates, usually at ambient temperature, by using PLD, sputtering or E-beam (or thermal) evaporation. The film thickness is typically 400-500 nm. This thickness is important to a certain extent because there is interdiffusion between films and some substrates above 600°C [44]. Then the precursors are sealed in an evacuated (or sometimes Ar-containing) Nb or Ta or quartz tube with Mg metal pieces. In some cases, the precursor films are wrapped in Nb or Ta envelopes with Mg metal pieces and this is all then encapsulated in an evacuated quartz tube. The amount of Mg has to provide an Mg vapor pressure sufficient to form $MgB_2$ at the annealing temperature. Nb or Ta tubes generally give better results than bare quartz tubes since Mg vapor has a detrimental effect on quartz tubes at elevated temperatures. Finally the tube is heated at 600-900°C for 10 to 60 min in an outside furnace (*ex-situ* annealing). This method is similar with the method that employed in high-quality Tl- or Hg-based superconducting cuprate films [45]. The annealing profile differs for each group. Several groups claim that rapid heating to the annealing temperature and quenching to room temperature may be important, while other groups do not. Table I summarizes the preparation recipes and physical properties of the two-step growth films reported by several groups. Early films prepared by two-step growth often contained a small quantity of MgO impurities, which can be significantly reduced by taking great care to avoid air exposure at every step of the process.

Figure 11 shows one of the best results for two-step growth reported by Kang *et al* [39]. The films are highly crystalline, and the superconducting transition temperature ($T_c$) is 39 K at zero resistance, which is the same as the bulk $T_c$. The residual resistivity ratio (RRR) is ~3. The RRR is only slightly smaller than the value of 5-6 reported for single crystals [46, 47]. As regards the critical current densities ($J_c$), the same group reported values of 40 MA/cm$^2$ at 5 K and 0 T, and ~0.1 MA/cm$^2$ at 15 K and 5 T [48] (Fig. 12). Furthermore low surface resistance ($R_s$) (19.4 μΩ at 4.2K and at 7.18 GHz) was obtained for these films [49]. Similar but slightly lower $J_c$ values were reported by Eom *et al.* and Moon *et al.* [40, 42]. The reported $J_c$ and $R_s$ values suggest that MgB$_2$ films are promising for practical applications.

Some groups have succeeded in preparing superconducting MgB$_2$ films by annealing precursor films "*in-situ*" in the growth chamber [43, 50-57]. With this process, the most serious problem is the limited Mg vapor pressure inside a vacuum chamber. So the annealing recipe is significantly different from that used in the "*ex-situ*" annealing process; the annealing temperature is lower, typically ~600°C, and the annealing time is shorter, typically a few to 20 min. Even with such quick low-temperature annealing, films often lose Mg, resulting in poor superconducting properties. To compensate for the Mg loss during *in-situ* annealing, precursor films are usually very rich in Mg. In some cases, even an extra Mg cap layer is deposited on the top of precursor films [53]. In other cases, films are exposed during annealing to Mg plasma that is generated by ablating Mg metal by PLD (Mg plasma annealing) [54].

4. 3 *In-situ* growth

In contrast to the case with two-step growth, only few groups have yet reported *in-situ* growth (as-grown films) of superconducting $MgB_2$ films [57-67]. Table II summarizes the preparation recipes and physical properties of the *in-situ* growth films reported by these groups.

Saito *et al.* prepared $MgB_2$ films by carrousel-type sputtering as shown in Fig. 13 [58, 67]. The Mg and B metal targets were placed on two adjacent cathodes, and each sputtering power was controlled independently. The carrousel with the substrate holder was rotated at 50 rpm during the deposition. The superconducting films were prepared at substrate temperatures ($T_S$) between 207ºC and 268ºC. The best $T_c$ of their films was ~28 K with a transition width of ~1 K.

Grassano *et al.* prepared $MgB_2$ films by PLD with an Mg enriched target (Mg:B = 1:1) at $T_S$ = 400-450ºC [59]. The crucial factor in obtaining superconducting films is that the growth should be undertaken in a blue laser plume (indicative of metallic Mg plasma) rather than a green plume (indicative of MgO plasma). This can be achieved only in a narrow range (~2 x $10^{-2}$ mbar) of Ar buffer gas pressure. The resultant films showed $T_c^{onset}$ ~ 25 K with a width of 2 K.

Three groups (Ueda *et al.* [57], Jo *et al.* [60], and Erven *et al.* [61]) reported the preparation of MgB$_2$ films by coevaporation. All three groups used a UHV chamber with a base pressure of 2-5 x 10$^{-10}$ Torr. The results obtained by the three groups are in good agreement with one another in most respects. The growth temperature needed to produce superconducting films is between 200°C and 300°C. A growth temperature higher than 300°C results in significant loss of Mg, and produces insulating films, even with an Mg rate 10 times the nominal rate. Below 300°C, too much Mg flux results in MgB$_2$ + Mg-solid, and too little Mg flux results in MgB$_2$ with Mg-deficit phases (MgB$_4$ etc.). All the groups failed to find a convenient "window" for MgB$_2$ + Mg-Gas. The resultant films showed $T_c$ = 33-35 K with a transition width of less than 1 K. The RRR value is 1.2-1.5. The microstructure of the films can be well illustrated by the plan-view TEM picture (Fig. 14) published by Jo *et al*. Hexagonal grains of ~400 Å in size are aligned rather closely but with 30° rotated coincidence. The films reported by the three groups, however, show some differences in resistivity value and crystallinity. The room-temperature resistivity value varies from 30 to 300 μΩcm. The rocking curve width (full width at half maximum) of the (002) X-ray diffraction peak varies from 0.7° to 3°, and the grain size determined by AFM varies from 20 to 50 nm.

Zeng *et al*. achieved the *in-situ* epitaxial growth of MgB$_2$ films by employing the hybrid physical-chemical vapor deposition (HPCVD) technique [62-66]. A schematic diagram of their equipment is shown in Fig. 15 [66]. They prepared the MgB$_2$ films in a CVD reactor in a diborane (B$_2$H$_6$), nitrogen and hydrogen gas flow. Mg was supplied from chips of bulk Mg that are located close to a substrate and heated together. The substrate and Mg chips are heated to above 700°C. The partial Mg pressure around the substrate becomes 0.1-1 Torr. The growth conditions with this method seem to lie in a "window" for $MgB_2$ + Mg-Gas. Excellent-quality single-crystalline $MgB_2$ films can be obtained with this method. As shown in Fig. 16, the films on *c*-cut SiC (either 4H-SiC with $a_0$ = 3.073Å or 6H-SiC with $a_0$ = 3.081Å) showed $T_c^{end}$ as high as 41.8 K, residual resistivity ($\rho_0$) as low as 0.28 $\mu\Omega$cm, and *RRR* above 30. Furthermore $J_c$ in a zero field reaches 3.5 x $10^7$ A/cm$^2$ at 4.2 K and $10^7$ A/cm$^2$ at 25 K. In reality, the superconducting properties of their films are even better than those of the best single crystals. With regard to multilayer deposition, however, it remains to be seen whether this method is as suitable as conventional physical vapor deposition methods.

4. 4. Substrates for $MgB_2$ films

The choice of substrate is important in terms of achieving less interdiffusion and better lattice matching. The substrates generally used for $MgB_2$ film fabrication are $Al_2O_3$-R, $Al_2O_3$-C, Si(100), Si(111), $SrTiO_3$(100), MgO(100), and SiC(0001). Since $MgB_2$ has a hexagonal $AlB_2$ structure, it should prefer substrates with a hexagonal face. The lattice matching between such substrates and $MgB_2$ is summarized in Table III.

As regards interdiffusion, He *et al.* examined the reactivity between $MgB_2$ and common substrate materials ($ZrO_2$, YSZ, MgO, $Al_2O_3$, $SiO_2$, $SrTiO_3$, TiN, TaN, AlN, Si and SiC) [44]. In their experiments, each of these substrate materials in fine powder form was mixed with Mg metal flakes and amorphous B powder, and reacted at elevated temperatures (600, 700 and 800ºC). Elemental Mg + B were employed in these reactions, rather than preformed $MgB_2$, to provide a better model of the film fabrication process. Furthermore very fine powder was used to enhance reactivity even at temperatures as low as ~600ºC, as often used in thin film preparation. Surprisingly, $MgB_2$ has been found to be rather inert to many substrate materials. Even at 800ºC, no reaction occurs with $ZrO_2$, MgO, or nitrides (TiN, TaN, AlN). The exceptions are $SiO_2$ and Si, where there is a severe reaction at 600ºC, and $Al_2O_3$, where a reaction is observed at 700ºC. At 800°C, $MgB_2$ is also reactive with $SrTiO_3$ and SiC. These results are helpful in terms of selecting appropriate substrates for thin film device applications.

**5. Growth of $MgB_2$ films at NTT**

5.1. Growth process

In this section, we describe our growth of $MgB_2$ films in detail [57]. $MgB_2$ thin films were grown by coevaporation in a custom-designed UHV chamber (base pressure ~5 x $10^{-10}$ Torr) from pure metal sources using multiple electron guns [68, 69]. This MBE chamber has been used for the growth of high-$T_c$ superconductor films for over a decade. The evaporation beam flux of Mg and B can be controlled by electron impact emission spectrometry (EIES) via feedback loops to the electron guns. The flux ratio of Mg to B was changed so that it was from 1 to 10 times as high as the nominal ratio. The growth rate was 1.5 - 2 Å/s, and the film thickness was 1000 Å for typical films. We used $Al_2O_3$-C, $Al_2O_3$-R, H-terminated Si(111), $SrTiO_3$(100), and glass (Corning #7059) substrates. We characterized the structure and crystallinity by X-ray diffraction (XRD) and reflection high-energy electron diffraction (RHEED). The composition was determined by inductively coupled plasma (ICP) spectrometry. We measured resistivity by the standard four-probe method using electrodes formed by Ag evaporation. The surface morphology and roughness of the films were characterized by AFM.

5.2 Growth temperature and Mg sticking coefficient

As described in Section 2, the thermodynamic and kinetic constraints for $MgB_2$ film growth predicts that the growth temperature should be well below 400°C with an Mg rate of around 10 Å/sec typical for vacuum deposition. Consistent with this prediction, we actually observed a significant Mg loss in films grown above a substrate temperature ($T_s$) of 300°C. Figure 17 shows the molar ratio of Mg to $B_2$ evaluated by ICP analysis in films grown at temperatures of 200°C to 500°C and also with Mg rates 1.3 to 10 times the nominal rate. The films grown above 300°C were significantly deficient in Mg even with a 10 times higher Mg rate. These Mg deficient films were transparent and insulating. This indicates that the upper limit for the growth temperature is 300°C. By approaching this upper limit, however, a slight and uncontrollable change (even ±5°C) in the growth temperature leads to a dramatic change in composition, and makes the results irreproducible. We have therefore chosen the following as our typical growth conditions: $T_s$ = 260-280°C with an Mg rate maintained at three times the nominal rate. The resultant as-grown films on $Al_2O_3$-C typically exhibited XRD patterns and ρ-T curves as shown in Fig. 18. The films had a $c$-axis preferred orientation, and their $T_c$ and resistivity were 32-35 K and 30-50 μΩcm at 300 K, respectively, which should be compared with the bulk single crystal values of 39 K and 5-10 μΩcm. Figure 19 shows an AFM image (1 μm×1 μm view) of one typical as-grown film. The grains are very small, typically 20-40 nm. The surface of the film is fairly smooth, and the root-mean-square roughness ($R_{MS}$) and the average roughness ($R_a$) are 2.2 nm and 1.8 nm, respectively.

When the growth temperature ($T_s$) is lowered, the crystallinity becomes poor judging from the XRD peak intensities, and simultaneously the superconducting properties become degraded. The $T_c$ value decreases monotonically as $T_s$ decreases and disappears below $T_s \sim 150°C$. The $T_c$ suppression is not only due to poorer crystallinity but also to excess Mg as described below.

5.3 Effect of excess Mg in films

Figure 20 shows the superconducting properties of $MgB_2$ films grown on $Al_2O_3$-C at a fixed substrate temperature (280°C) but with different Mg rates (1-10 times). The desired phase is not formed with Mg rates of less than 3 times the nominal rate. At higher Mg rates, $T_c$ gradually decreases and eventually becomes around 15 K at 10 times the nominal Mg rate. A reduction in resistivity accompanies this $T_c$ suppression, indicating the presence of metal Mg in these films. This result demonstrates the harmful effect of excess Mg on the superconducting properties, which is either due to the proximity effect with normal Mg metal or the formation of nonstoichiometric $Mg_{1+x}B_2$ that hitherto was not known to be formed in bulk synthesis.

5.4 Effect of residual oxygen during growth

In our early attempts to grow $MgB_2$, we had some difficulty in obtaining reproducible results. This can be partly explained by the fact that the Mg sticking coefficient varies dramatically near $T_s \sim 300°C$. However, we have noticed another reason for the irreproducibility, namely the effect of residual oxygen on $MgB_2$ film growth. Hence we undertook a systematic investigation of the effect of residual oxygen on the growth of $MgB_2$ films. We varied the oxygen partial pressure from < 1

$\times 10^{-10}$ to $8\times 10^{-6}$ Torr.

Figure 21 shows the XRD patterns and ρ-T curves of MgB$_2$ films grown on Al$_2$O$_3$-C under different partial oxygen pressures. Weak but definite MgB$_2$ (00$l$) peaks are observed for the film grown in $P_{O2}$ < $1.0\times 10^{-10}$ Torr whereas no peak is observed for the film grown in $P_{O2}$ ~ $3.7\times 10^{-9}$ Torr. The resistivity of the latter film is much higher although $T_c$ is not greatly suppressed (only ~3 K). For $P_{O2} \geq 10^{-7}$ Torr, the films were transparent and insulating, indicating that no MgB$_2$ phase was formed. These observations strongly indicate that residual oxygen is very harmful to MgB$_2$ film growth, even with $P_{O2}$ as low as $1\times 10^{-9}$ Torr. This result can be readily understood from the high sensitivity of Mg to oxidation. Exactly the same conclusion was reached by Erven *et al*.

5.5 Effect of *in-situ* annealing

As seen from the above results, the properties of our as-grown films are inferior to those of bulk single crystals or *ex-situ* post-annealed films or as-grown epitaxial films obtained by HPCVD. The superconducting transition temperature is lower, namely $T_c^{end}$ is ~35 K at its highest, and the resistivity has a higher value and a weaker temperature dependence. With the aim of improving the superconducting properties of MgB$_2$ films, we undertook *in-situ* post-annealing for as-grown films. Our *in-situ* post-annealing was performed just after growth at annealing temperatures ($T_a$) of 380-680 °C for 10 minutes while exposing the films to Mg flux (~4.5 Å/sec). From the thermodynamic viewpoint, our *in-situ* post-annealing may be performed outside the stability field of the MgB$_2$ phase. As described in Sec. 2.3, however, the very slow kinetics of the decomposition process predicts that the inside of the films will be protected from decomposition (see Fig. 4) [7]. In fact, this *in-situ* post-annealing slightly improves the inside of the films as shown in Fig. 22. The superconducting transition temperature increased gradually with increasing $T_a$, and a maximum improvement of ~2 K was achieved by annealing at $T_a = 503°C$. The resistivity was also improved although there was some scattering in the data. This improved superconductivity may result from a partial elimination of stress at the grain boundaries. This is because there is almost no improvement in the crystallinity of the films by annealing, judging from the XRD patterns. The RHEED observations during annealing indicated that the surface of the films starts to decompose at $T_a \sim 500°C$. The RHEED patterns became halo-like with annealing at $T_a \sim 550°C$, above which the inside of the films also started to decompose. This situation did not change greatly even when the Mg flux rate was tripled to ~15.0 Å/sec.

5.6 Thickness dependence

The thickness dependence of the superconductivity was examined for both as-grown and post-annealed films. We undertook this investigation specifically to evaluate the degraded surface thickness of post-annealed films. The film thickness, which was controlled by changing the deposition time, was varied from 2.5 to 100 nm. In this experiment, the post annealing temperature was fixed at 480°C. The thickness dependences of the ρ-*T* curves are shown for as-grown films in Fig. 23 (a), and for post-annealed films in Fig. 23 (b). In both cases, $T_c$ decreased and the resistivity increased with decreasing film thickness. Figure 24 is a plot of $T_c$ as a function of film thickness. For film thicknesses greater than 10 nm, the $T_c$ of the post-annealed films is higher than that of the as-grown films, but the situation is reversed at 5 nm. The as-grown film, even with a thickness of 5 nm, is superconducting with $T_c^{end}$ of ~10 K, whereas the post-annealed 5 nm film is insulating. This implies that a surface with a thickness of more than 5 nm decomposes due to Mg loss even with $T_a = 480°C$. Our recent tunneling experiments support this conclusion: post-annealed films have a dead surface (nonsuperconducting) layer and are unsuitable for fabricating superconducting junctions [70].

5.7 Effect of substrates

Finally we describe the effect of the substrates. Since our as-grown films are not highly crystalline, the epitaxial effect of the substrates is relatively unimportant as compared with high-$T_c$ cuprate films, where the substrates play a significant role in improving the film quality [71]. However, we observed that $MgB_2$ films have a weak but finite substrate dependence. Figure 25 shows the superconducting transitions of $MgB_2$ films on different substrates: (a) as-grown and (b) post-annealed at $T_a = 480°C$. For the as-grown films, $T_c^{zero}$ was 32.2, 33.2, 30.3 and 29.5 K for Si(111), $Al_2O_3$-C, $Al_2O_3$-R, and $SrTiO_3$(100), respectively. For the post-annealed films, $T_c^{zero}$ was 36.8, 36.6, 36.1 and 35.7 K for Si(111), $Al_2O_3$-C, $Al_2O_3$-R, and $SrTiO_3$(100), respectively. In both cases, $T_c^{zero}$ was slightly higher on hexagonal Si(111) and $Al_2O_3$-C than on rectangular $Al_2O_3$-R or square $SrTiO_3$(100).

We also prepared $MgB_2$ films on glass substrates (Corning #7059). The as-grown film showed a $T_c^{zero}$ of 33.4 K. This result demonstrates that fair quality $MgB_2$ films can be grown even on amorphous substrates such as glass.

**6. $MgB_2$ junctions**

Superconducting junctions are important in terms of electronic applications as well as basic research. $MgB_2$ may be suitable for fabricating junctions because it has less anisotropy, fewer material complexities, and a longer coherence length ($\xi$ = ~5 nm) than high-$T_c$ cuprates. Various types of superconducting $MgB_2$ junctions have already been examined including point-contact or break junctions [72-74], nanobridges [75], planar junctions by localized ion damage [76], ramp-type junctions [77], sandwich-type junctions [78, 79], etc. In this section, we describe some typical results.

6.1 Point-contact junctions

Zhang *et al.* employed a point-contact method using two pieces of $MgB_2$, and obtained either SIS or SNS junctions by adjusting the contact pressure [72]. With loose contact, they obtained SIS characteristics that show fairly good quasi-particle spectra with an energy gap of 2.0 meV as shown in Fig. 26. With tight contact, they obtained SNS characteristics that fit the prediction of the resistively-shunted junction (RSJ) model. The dc SQUIDs made from two SNS junctions yielded magnetic flux noise and field noise as low as 4 $\mu\Phi_0$ $Hz^{-1/2}$ and 35 fT $Hz^{-1/2}$ at 19 K, where $\Phi_0$ is the flux quantum. The low-frequency noise is 2 - 3 orders of magnitude lower than that of the YBCO SQUID early in its development, indicating that $MgB_2$ has excellent potential for providing a SQUID that operates around the temperature of 20-30 K reached by current commercial cryocoolers.

6.2 Nanobridges

Brinkman *et al.* fabricated SQUIDs using nanobridges (~70 nm wide, ~150 nm high, and ~150 nm long) patterned by focused-ion-beam (FIB) milling on $MgB_2$ films with $T_c$ = 24 K prepared by a two-step *in-situ* process [75]. These bridges have outstanding critical current densities of 7 x $10^6$ A/cm$^2$ at 4.2 K, indicating that the superconductivity of MgB$_2$ survives by nanostructuring even at a length scale < 100 nm, which is in complete contrast to high-$T_c$ cuprates. The device operated below 22 K, and a modulation voltage of 30 µV was observed at 10 K.

Burnell *et al*. fabricated planar junctions with localized ion damage and Ga implantation in MgB$_2$ thin films with $T_c$ = 36 K prepared by a two-step process [76]. The barrier was defined by writing cuts of width $d$ = 50 nm across the track width using a 4pA 30 kV Ga ion beam. The junctions showed nonhysteretic RSJ-like characteristics with rather high $I_cR_n$ (~1 mV at 4 K). The critical current of their junctions was strongly modulated by applied microwave radiation and magnetic field.

6.3 Ramp-type junctions

(1) Mijatovic *et al*. fabricated ramp-type junctions [77]. Their junction fabrication process is very similar to the process used for ramp-edge junctions for high-$T_c$ cuprates. The bottom and top MgB$_2$ layers were prepared by a two-step *in-situ* process, and separated by a 12-nm MgO barrier layer. The junctions showed RSJ-like current-voltage characteristics and both dc and ac Josephson effects were observed. The $I_cR_n$ product was ~0.13 mV at 4 K, which is much smaller than expected from the energy gap of MgB$_2$ or $I_cR_n$ reported in point-contact or FIB damaged junctions.

6.4 Sandwich-type junctions

As yet there has been no successful report on fabricating sandwich-type

MgB$_2$/I/MgB$_2$ Josephson tunnel junctions, but SIS' or SIN junctions with one MgB$_2$ electrode and one low-$T_c$ superconductor or normal metal electrode have been reported by a few groups [78, 79].

Saito *et al*. fabricated sandwich-type NbN/AlN/MgB$_2$ junctions [79]. The bottom MgB$_2$ layers ($T_c$ ~28 K) were prepared in an *in-situ* process by carrousel-type sputtering, which is described in §4.2. Subsequently, in the same chamber, 0.8–2.0 nm thick AlN barrier layers and 50 nm thick Nb top layers were deposited by reactive sputtering. The junctions showed fair characteristics (Fig. 27(a) (b)) with a substantial Josephson supercurrent ($I_cR_n$ ~ 0.9 mV), a clear gap edge, and fairly small subgap conductance. Furthermore, an almost ideal Fraunhofer pattern was observed as shown in Fig. 27(c), indicating uniform tunneling current in the junction. The superconducting gap value for MgB$_2$ was estimated to be 2.95 meV, which, they claimed, is in fair agreement with the BCS prediction with $T_c^{end}$ ~ 23K.

We also fabricated sandwich-type SIN junctions on as-grown MgB$_2$ films prepared by coevaporation ($T_c$ ~ 33 K) [70, 80]. We have tried different barrier layers and also different normal metals for the top electrode. The best combination so far obtained is Au/MgO/MgB$_2$. In these junctions, thin Mg metal (1-3 nm) was deposited on MgB$_2$ films, and subsequently oxidized in air. The junctions showed a reproducible and well-defined superconducting gap ($\Delta$ = 2.0-2.5 meV) as shown in fig. 28. The $2\Delta/k_BT_c$ = 1.4-1.8 is significantly smaller than the BCS value, indicating that the observed gap corresponds to the smaller gap ($\Delta_s$) in the multi-gap scenario.

## 6. Conclusion

In this review article, we made a broad survey of the efforts to the growth of superconducting $MgB_2$ films made in the first 30 months since the discovery together with a small review of chemistry and physics of $MgB_2$. We also gave a brief summary of the past efforts at fabricating $MgB_2$ junctions. The following is the summary of this article.

Chemistry of $MgB_2$

Magnesium diboride, $MgB_2$, has a hexagonal $AlB_2$ structure. The structure consists of hexagonal-close-packed (hcp) layers of Mg atoms alternating with graphite-like honeycomb layers of B atoms. The lattice constants are a = 3.086 Å and c = 3.524 Å.

In the Mg-B system, there are three intermediate compounds, $MgB_2$, $MgB_4$, and $MgB_7$, in addition to the gas, liquid, and solid magnesium phases and the solid boron phase. $P_{Mg}$ has a significant influence on the decomposition temperature ($T_d$): $T_d$ = 1545°C at 1 atm, 912°C at 1 Torr, and 603°C at 1 mTorr. Above $T_d$, $MgB_2$ decomposes into a mixture of $MgB_4$ and Mg vapor.

The vapor pressure curve of elemental Mg is significantly higher than that of $MgB_2$. Below 400°C, the reaction kinetics is slow, therefore Mg reevaporates from the surface before Mg is trapped by B, resulting in no $MgB_2$ phase formation.

The decomposition reaction of $MgB_2$ is very slow because of a large kinetic barrier. The evaporation coefficient ($\alpha$), which is defined as the ratio of the observed Mg rate to the thermodynamically predicted rate, is $10^{-4}$.

Physics of $MgB_2$

MgB$_2$ appears to be a conventional, BCS phonon-mediated *sp* superconductor. The $T_c$ as high as 40 K can be achieved by high phonon frequencies due to light mass elements composing this compound. The Debye temperature for MgB$_2$ is over 1000 K. The specific boron in-plane ($E_{2g}$) vibration mode bond-stretching mode) is strongly coupled to electrons (B $2p\sigma$ band) near the Fermi level. The electron-phonon coupling $\lambda$ can be calculated as 0.9-1.0 by a deformation potential approximation, indicating an intermediate coupling regime.

MgB$_2$ is a two-gap superconductor. The larger gap $\Delta_l$ is ~6 meV ($2\Delta_l/k_B T_c \sim 4$) and the smaller gap $\Delta_S$ is ~2 meV ($2\Delta_S/k_B T_c \sim 1.2$). The postulation of the two-gap superconductivity of MgB$_2$ can be justified from the band calculations. The bands near the Fermi level mainly derive from two distinct sets of boron orbitals: $sp^2$ ($\sigma$) states and $p_z$ ($\pi$) states. The $\sigma$ bands are 2D in character and responsible for the larger superconducting gap, whereas the $\pi$ bands have more of a 3D character and are responsible for the smaller gap.

Most of the physical properties are relatively normal, which is in contrast to those of high-$T_c$ cuprates. The resistivity can be well fitted by the Bloch-Grüneisen formula with a very high Debye temperature of 1000 ± 100 K [26]. The resistivity value of MgB$_2$ is 5-6 μΩcm at 300 K and 0.3 μΩcm to 3 μΩcm just above $T_c$. A rough estimate of the carrier density from $R_H$ (*H*//*c*) gives 2-3 x 10$^{23}$ cm$^{-3}$, which is similar to that of conventional metals and two orders-of-magnitude larger than that of YBCO.

The upper critical field $H_{c2}^{//c}$ is low, about 3 to 4 T, even at $T = 0$ K, and $H_{c2}^{//ab}$ reaches 15 to 20 T at $T = 0$ K. The resultant anisotropy ($H_{c2}^{//c} / H_{c2}^{//ab}$) is 4-5. The GL coherence length is given as $\xi_{ab} = 10\pm2$ nm, $\xi_c = 2$-3 nm. The mean free path ($l_{ab}$) is as large as 50-100 nm [34], which indicates that MgB$_2$ is in the clean limit. The lower critical field $H_{c1}$ is also anisotropic and ~30 mT. The magnetic penetration depth is given as $\lambda_{ab}$ = 110-130 nm and $\lambda_c$ = 210-280 nm.

MgB$_2$ film growth

(1) The most serious problem in MgB$_2$ film growth is the high vapor pressure of Mg required for phase stability. There are two ways to overcome it. One is to prepare the film under a high Mg vapor pressure in a confined container at a high temperature ("two-step growth"), and the other is to prepare the film at a low temperature ("in-situ growth"). Two-step growth produces good crystalline films with superconducting properties comparable to those of high-quality single crystals although it cannot be used to fabricate Josephson junctions or multilayers. By contrast, *in-situ* growth can produce only poor crystalline films that have a slightly lower T$_c$ (typically 35 K) than the bulk value, but this approach makes multilayer deposition feasible.

(2) Best films by two-step growth have $T_c$ ~ 39 K at zero resistance, $\rho$(300K) ~ 10 $\mu\Omega$cm, and RRR is ~3. The critical current densities ($J_c$) is as high as 40 MA/cm$^2$.

(3) *In-situ* PVD growth is performed by sputtering, PLD, and MBE. The best films have so far been obtained by MBE. The best films have have $T_c$ ~ 35 K at zero resistance, $\rho$(300K) ~ 30 $\mu\Omega$cm, and *RRR* is ~1.5. A growth temperature higher than 300°C results in significant loss of Mg, and produces insulating films. even with an Mg rate 10 times the nominal rate. Below 300°C, too much Mg flux results in MgB$_2$ + Mg-solid, and too little Mg flux results in MgB$_2$ with Mg-deficit phases (MgB$_4$ etc.). No group could find a convenient "window" for MgB$_2$ +

Mg-Gas.

(4) "Hybrid physical-chemical vapor deposition (HPCVD) technique produced the highest-quality single-crystalline $MgB_2$ films. The films have $T_c^{end}$ as high as 41.8 K, residual resistivity ($\rho_0$) as low as 0.28 $\mu\Omega$cm, and *RRR* above 30. Furthermore $J_c$ in a zero field reaches 3.5 x $10^7$ A/cm$^2$ at 4.2 K and $10^7$ A/cm$^2$ at 25 K. The superconducting properties of their films are even better than those of the best single crystals.

(5) The substrates generally used for $MgB_2$ film fabrication are $Al_2O_3$-R, $Al_2O_3$-C, Si(100), Si(111), $SrTiO_3$(100), MgO(100), and SiC(0001). Since $MgB_2$ has a hexagonal $AlB_2$ structure, it should prefer substrates with a hexagonal face. $MgB_2$ has been found to be rather inert to many substrate materials. Even at 800ºC, no reaction occurs with $ZrO_2$, MgO, or nitrides (TiN, TaN, AlN). The exceptions are $SiO_2$ and Si, where there is a severe reaction at 600ºC, and $Al_2O_3$, where a reaction is observed at 700ºC.

<u>$MgB_2$ junctions</u>

(1) Various types of superconducting $MgB_2$ junctions have already been examined. Point-contact method using two pieces of $MgB_2$ showed good quasi-particle spectra with an energy gap of 2.0 meV with loose contact and SNS characteristics that fit the prediction of the resistively-shunted junction (RSJ) model with tight contact. The dc SQUIDs made from two SNS junctions yielded magnetic flux noise and field noise as low as 4 $\mu\Phi_0$ Hz$^{-1/2}$ and 35 fT Hz$^{-1/2}$ at 19 K.

(2) Nanobridges (~70 nm wide, ~150 nm high, and ~150 nm long) showed outstanding critical current densities of 7 x $10^6$ A/cm$^2$ at 4.2 K, indicating that the superconductivity of $MgB_2$ survives by nanofabrication even at a length scale < 100 nm, which is in complete contrast to high-$T_c$ cuprates.

(3) As yet, there has been no successful report on fabricating sandwich-type $MgB_2$/I/$MgB_2$ Josephson tunnel junctions, but SIS' or SIN junctions with one $MgB_2$ electrode and one low-$T_c$ superconductor or normal metal electrode showed promising results, in contrast to high-$T_c$ cuprates. NbN/AlN/$MgB_2$ junctions showed fair characteristics with a substantial Josephson supercurrent ($I_cR_n$ ~ 0.9 mV), a clear gap edge, and fairly small subgap conductance. Furthermore, an almost ideal Fraunhofer pattern was observed. SIN junctions showed a reproducible and well-defined superconducting gap ($\Delta$ = 2.0-2.5 meV). The $2\Delta/k_BT_c$ = 1.4-1.8 indicates that the observed gap corresponds to the smaller gap ($\Delta_s$) in the multi-gap scenario.

Future prospects

Finally our comment are given on the prospect of $MgB_2$ for superconducting digital applications in comparison with high-$T_c$ cuprates. In the case of high-$T_c$ cuprates, after 17 years since the discovery, no reliable fabrication method of good Josephson junctions is yet established. With regard to this unsuccess, we pointed out an intrinsic and serious problem of cuprates [81]. That is redox reaction at interface between cuprates and other materials. Oxygen in cuprates is loosely bound and easily taken out by contact with other materials. It leads to serious degradation of superconductivity at interface of cuprates. Without precise control of interface oxygen, it is very hard to obtain reproducible results.

In contrast, in the case of $MgB_2$, fair quality of SIS' and SIN junctions have already been obtained, and the results are in fair agreement.  This seems to indicate that there may be no complicating interface problem in MgB$_2$.  Our preliminary attempts to fabricate SIS junctions indicated that the problem is mainly leaky barrier.  This problem may be overcome by careful selection of barrier materials.  So, we believe that MgB$_2$ may be promising for superconducting electronics applications in spite of $T_c$ not as high as high-$T_c$ cuprates.

We hope this review would help the efforts toward superconducting electronics applications using MgB$_2$.  Finally we apologize that many fresh important reports could not be included in this article mostly because of our insufficient capacity.

Figure captions

Fig. 1.   Crystal structure of $MgB_2$.

Fig. 2.   (Reproduced from ref. [7]) Temperature-composition phase diagram of the Mg-B system under the Mg pressures of (a) 1 atm, (2) 1 Torr, and (c) 1 mTorr.

Fig. 3. (Reproduced from ref. [7]) Pressure-temperature phase diagram for the Mg : B atomic ratio $x_{Mg} / x_B \geq 1/2$.   The window of "$MgB_2$ + Mg-gas", which is convenient for the growth of $MgB_2$ films, can be approximately expressed by the following equations: $\log(P) = -7561/T + 8.673$ (upper boundary) and $\log(P) = -10142/T + 8.562$ (lower boundary).

Fig. 4.   Comparison of the Mg vapor pressure over Mg metal (a) and $MgB_2$ (b).   The Mg pressure over Mg metal is significantly higher than that over $MgB_2$.   The kinetically limited Mg evaporation pressure (c) [8] is also shown, which is well below the thermodynamic decomposition curve.

Fig. 5.   (Reproduced from ref. [12]) (a) Generalized phonon density of states (PDOS) for $MgB_2$ determined from inelastic neutron scattering measurements (top curve) and the calculated PDOS (bottom curve).   (b) Calculated phonon dispersion curves and corresponding DOS (right panel).   At the zone center ($\Gamma$), the symmetries of the modes are also indicated.   (c) $E_{2g}$ modes that dominate the electron-phonon coupling.

Fig. 6. (Reproduced from ref. [18]) Experimental specific heat data (o) as a function of the reduced temperature $t$ (=$T/T_c$) from two groups [17, 29]. The data are compared with the BCS-normalized specific heat (thin line) and also two-gap fits (thick line). Insets: gaps $2\Delta_L/k_BT_c$ and $2\Delta_S/k_BT_c$ versus $t$ (dotted lines) and partial specific heat of both bands (full lines).

Fig. 7. (Reproduced from ref. [19]) (a) Superfluid density [$\rho = \lambda^2(0) / \lambda^2(T)$] of the polycrystalline sample (o), along with a fit to the two-gap model (solid line). The contributions of the small gap ($\Delta_S$) and the large gap ($\Delta_L$) in the model are also shown. (b) Superfluid densities of the single crystal sample ($\rho_{ab}$ (o) and $\rho_e$ ($\Delta$), which almost coincide with each other), along with a fit to the two-gap model (solid line). The $\rho_{ab}$ is calculated from $\lambda_{ab}$, whereas the $\rho_e$ is calculated from the H $\parallel$ ab data, which is a mixture of $\lambda_c$ and $\lambda_{ab}$. The predicted behavior of $\rho_{ab}$ and $\rho_c$ in the anisotropic gap model ($\Delta(\cos(\theta) = \Delta_0(1 + a\cos^2(\theta))/(1 + a)$) with a = 2.2 is shown by the dashed lines (denoted as $\rho_{ab}^{AG}$ and $\rho_c^{AG}$).

Fig. 8. (Reproduced from ref. [26]) Resistivity of an $MgB_2$ single crystal as a function of temperature. The residual resistivity ratio is about 5. The solid line is a fitting curve obtained using the Bloch-Grüneisen formula with a Debye temperature ($\Theta_D$) ~1100 K.

Fig. 9. (Reproduced from ref. [28]) The in- and out-of-plane Hall constants, as a function of temperature in the normal state of $MgB_2$ single crystals (top and bottom panels, respectively).

Fig. 10. (Reproduced from ref. [32]) Temperature dependence of the anisotropic upper critical fields obtained by transport (circles), ac-susceptibility (squares), and specific-heat (triangles) measurements for $MgB_2$ single crystals. The inset shows the temperature dependence of the anisotropy.

Fig. 11. (Reproduced from ref. [39]) (a) XRD patterns of $MgB_2$ films prepared by two-step growth on $SrTiO_3$ (A) and $Al_2O_3$-R (B) substrates. (b) Resistivity vs temperature in $H = 0$ and 5 T of an $MgB_2$ film prepared by two-step growth on $Al_2O_3$-R substrates. The lower inset is an enlarged view near $T_c$.

Fig. 12. (Reproduced from ref. [38, 48]) (a) Temperature dependence of the critical current density of $MgB_2$ films prepared by two-step growth on $Al_2O_3$-R substrates in H = 0 – 5 T evaluated from *M-H* curves (open symbols) and *I-V* curves (solid symbols). (b) Temperature dependence of the surface resistance ($R_s$) of an $MgB_2$ film measured at 7.18 GHz.

Fig. 13. (Reproduced from ref. [67]) Schematic picture of the carrousel sputtering employed by Saito *et al*.

Fig. 14. (Reproduced from ref. [60]) Plan-view TEM and diffraction pattern of the $MgB_2$ film grown *in-situ* by coevaporation.

Fig. 15. (Reproduced from ref. [66]) Schematic diagram of the HPCVD reactor employed by Zeng *et al*. A substrate with Mg chips is placed on a plate that can be heated with external induction coils. At 700 °C, none of the Mg sticks to the substrate until a small amount in diborane ($B_2H_6$) is added to the nitrogen + hydrogen gas flow, and stoichiometric $MgB_2$ begins to deposit on the substrate.

Fig. 16. (Reproduced from ref. [65]) Resistivity versus temperature for a 2250-Å-thick $MgB_2$ film grown on 4H-SiC by HPCVD. The inset shows an enlarged view near $T_c$.

Fig. 17. Molar ratio of Mg to $B_2$ evaluated by ICP analysis in films grown at 200°C to 500°C with Mg flux rates 1.3 to 10 times the nominal rate. The Mg loss is significant above $T_s = 300$°C.

Fig. 18. (a) XRD pattern and (b) $\rho$-$T$ curve of a typical "as-grown" $MgB_2$ film grown on an $Al_2O_3$-C substrate. The insets in (a) and (b) show a RHEED pattern and an enlarged view of the superconducting transition for the film.

Fig. 19. AFM image (1 μm x 1 μm) of an as-grown $MgB_2$ film.

Fig. 20. (a) $T_c$ vs Mg flux ratio and (b) $\rho$-$T$ curves of $MgB_2$ films grown with different Mg fluxes at a fixed substrate temperature (280°C).

Fig. 21. (a) XRD patterns and (b) $\rho$-$T$ curves of $MgB_2$ films grown under different $O_2$ pressures. The inset in (b) is an enlarged view of the superconducting transition.

Fig. 22. Superconducting transitions of $MgB_2$ films annealed at different temperatures.

A maximum improvement of ~2 K was achieved at $T_a$ = 503°C.

Fig. 23. Resistivity-temperature curves of $MgB_2$ films with different thicknesses: (a) as-grown films and (b) *in-situ* post-annealed films (480°C for 10 min).

Fig. 24. Thickness dependence of $T_c$ of as-grown films and *in-situ* post-annealed films (480°C for 10 min).

Fig. 25. Resistivity –versus temperature curves of (a) as-grown and (b) *in-situ* post-annealed films (480°C for 10 min) around $T_c$ prepared on various substrates (Si (111) (●), sapphire C (○), sapphire R (□), $SrTiO_3$ (100) (◇)).

Fig. 26. (Reproduced from ref. [72]) Current (crosses) and conductance dI/dV (diamonds) vs. voltage for $MgB_2$ tunnel junctions fitted to the theory shown as solid and dotted curves. Temperature is (a) 8.9 K and (b) 16.4 K. The inset in (a) is a Δ vs. temperature curve.

Fig. 27. (Reproduced from ref. [79]) Typical I-V and dI/dV-V characteristics for two $MgB_2$/AlN/NbN tunnel junctions with 20×20 μm junctions areas that was measured at 4.2 K. The AlN thickness is (a) 2.0 and (b) 0.8 nm, respectively. (c) Magnetic-field dependence of critical supercurrents ($I_c$) for a 20×20μm junction with $I_c$= 125 μA. The vertical axis is the current scale in 20-μA divisions, and the horizontal axis is the dc magnetic field on an arbitrary scale.

Fig. 28. Temperature dependence (1.4 - 41.1 K) of the tunneling spectra of the Au/MgO(Mg:10Å)/MgB$_2$ tunnel junction.

Table I. Recipes for film preparation and the physical properties of films prepared by the two-step growth technique.

Table II. Recipes for film preparation and the physical properties of resultant films prepared by as-grown synthesis.

Table III. The crystal structure and lattice constants of MgB$_2$ and several widely used substrates.

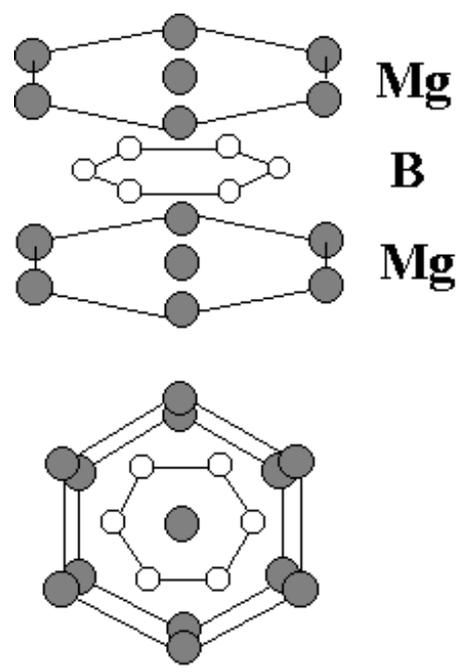

**Fig. 1**

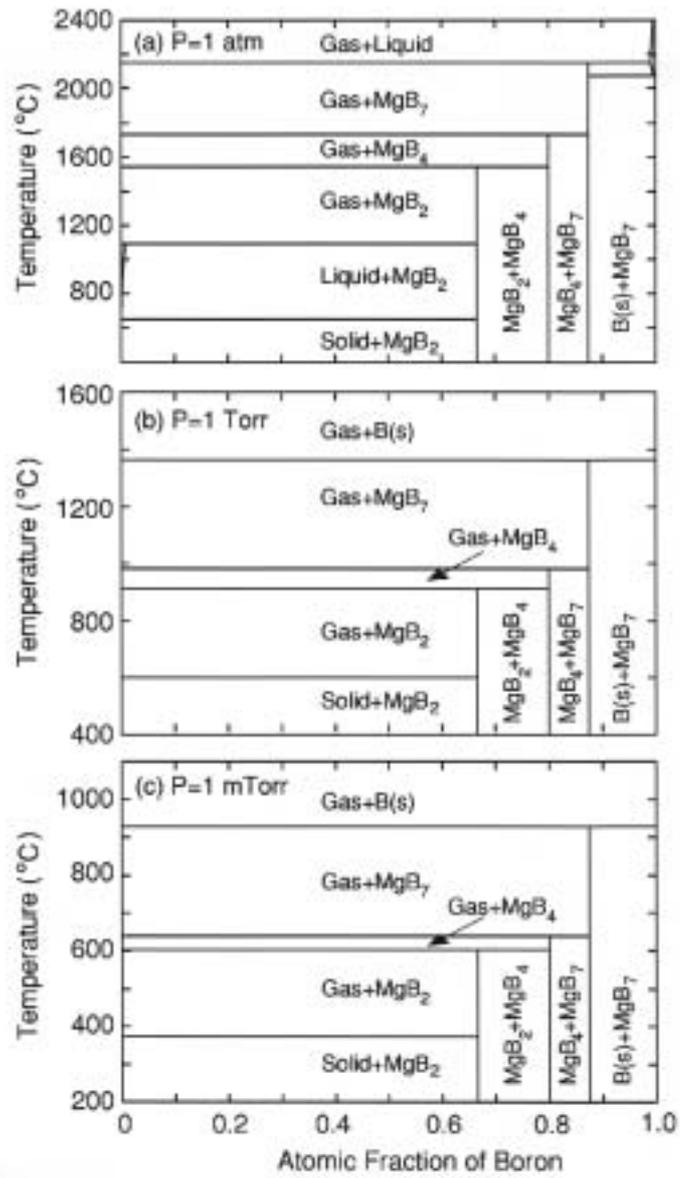

**Fig. 2**

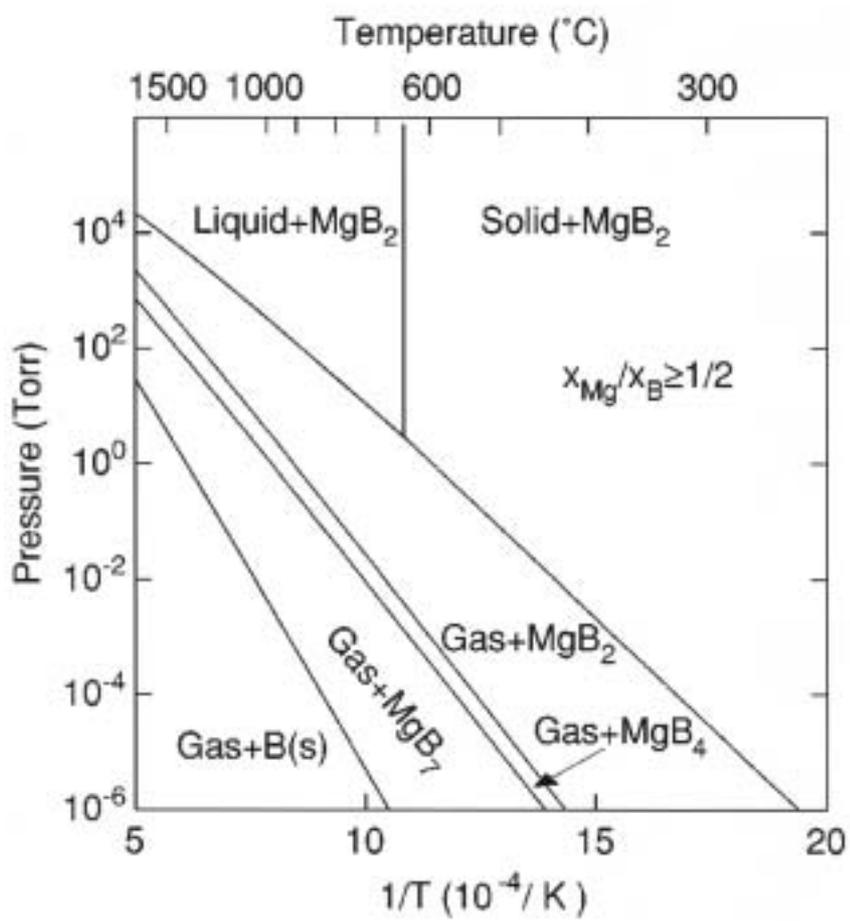

**Fig. 3**

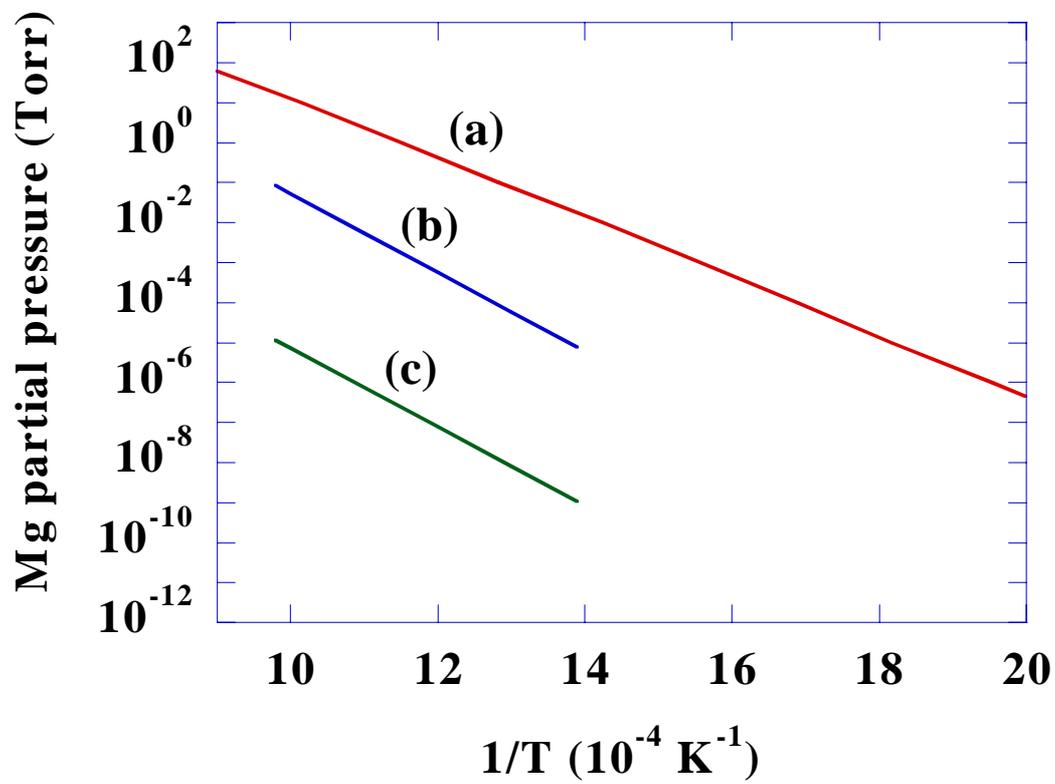

Fig. 4

**(a)**

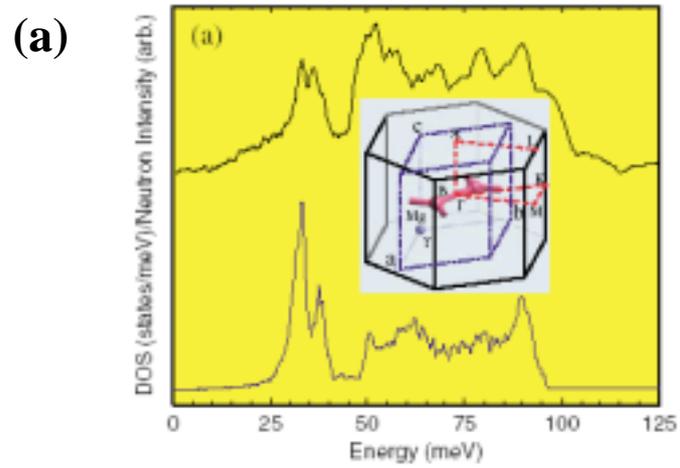

**(b)**

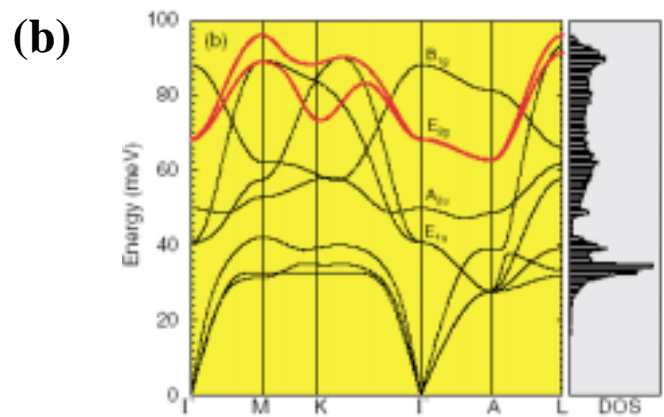

**(c)**

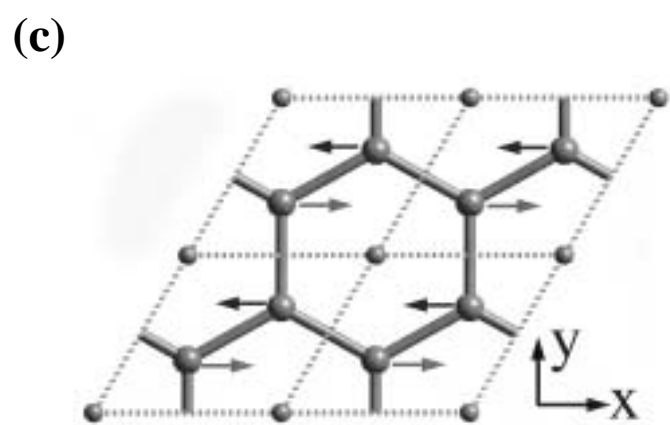

**Fig. 5**

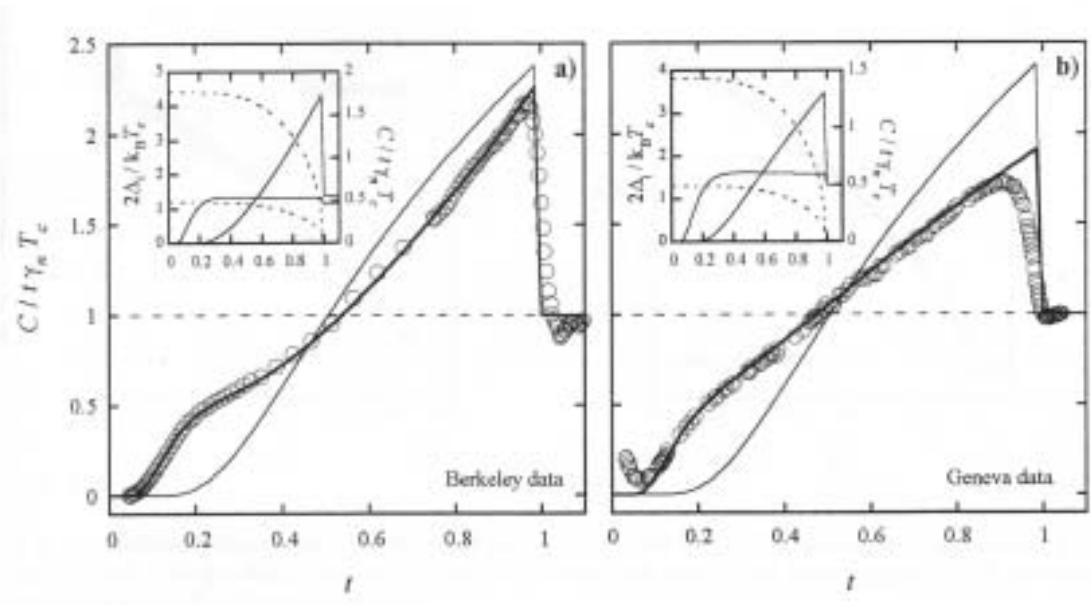

**Fig. 6**

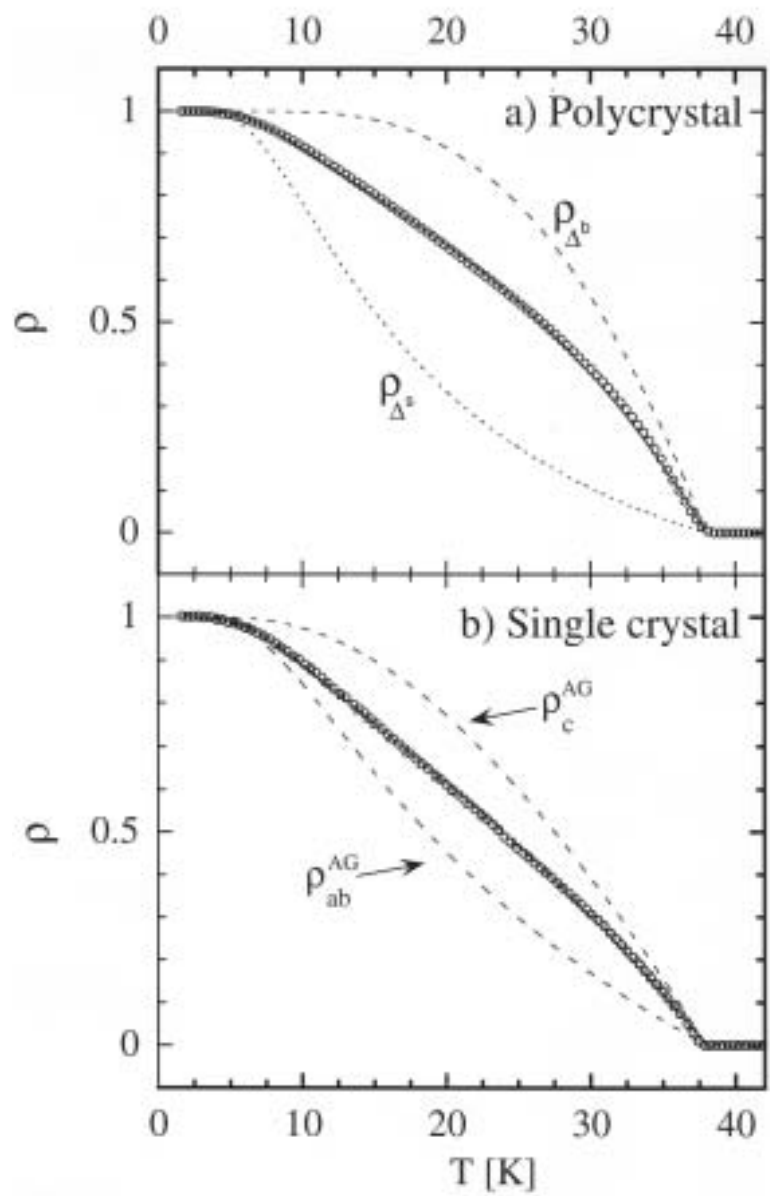

**Fig. 7**

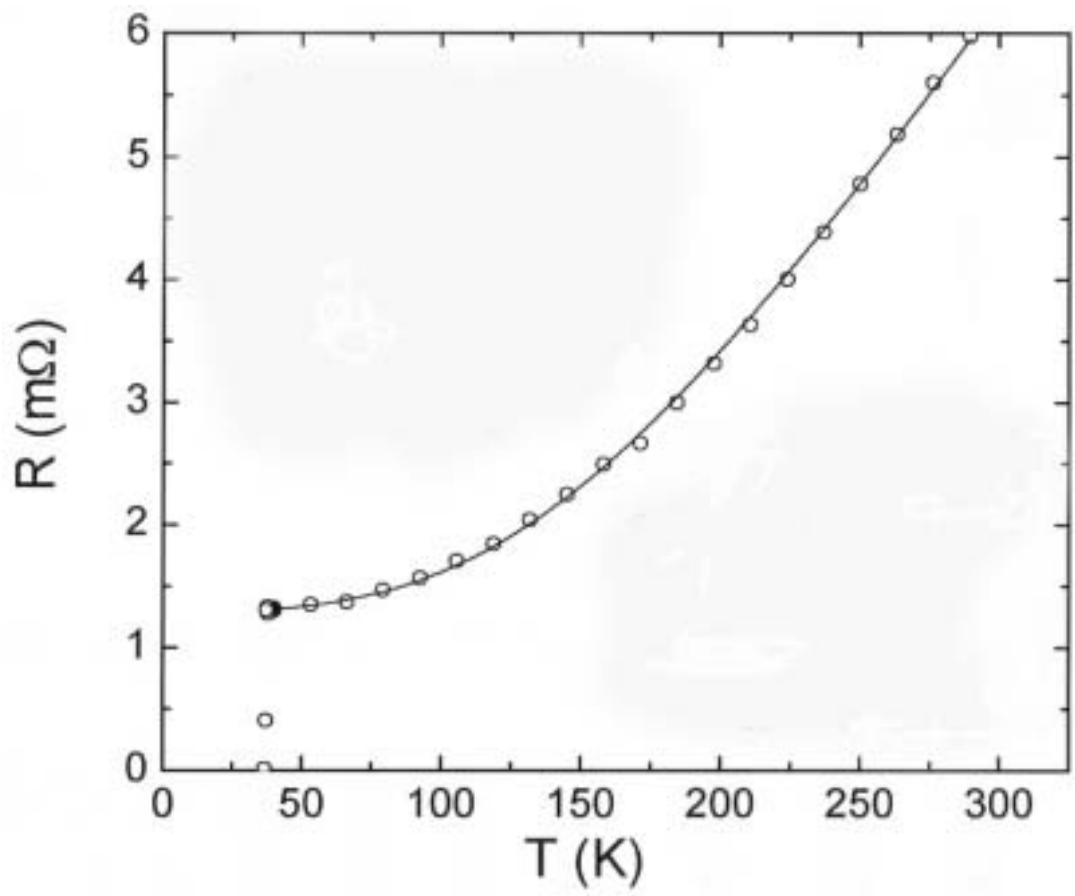

Fig. 8

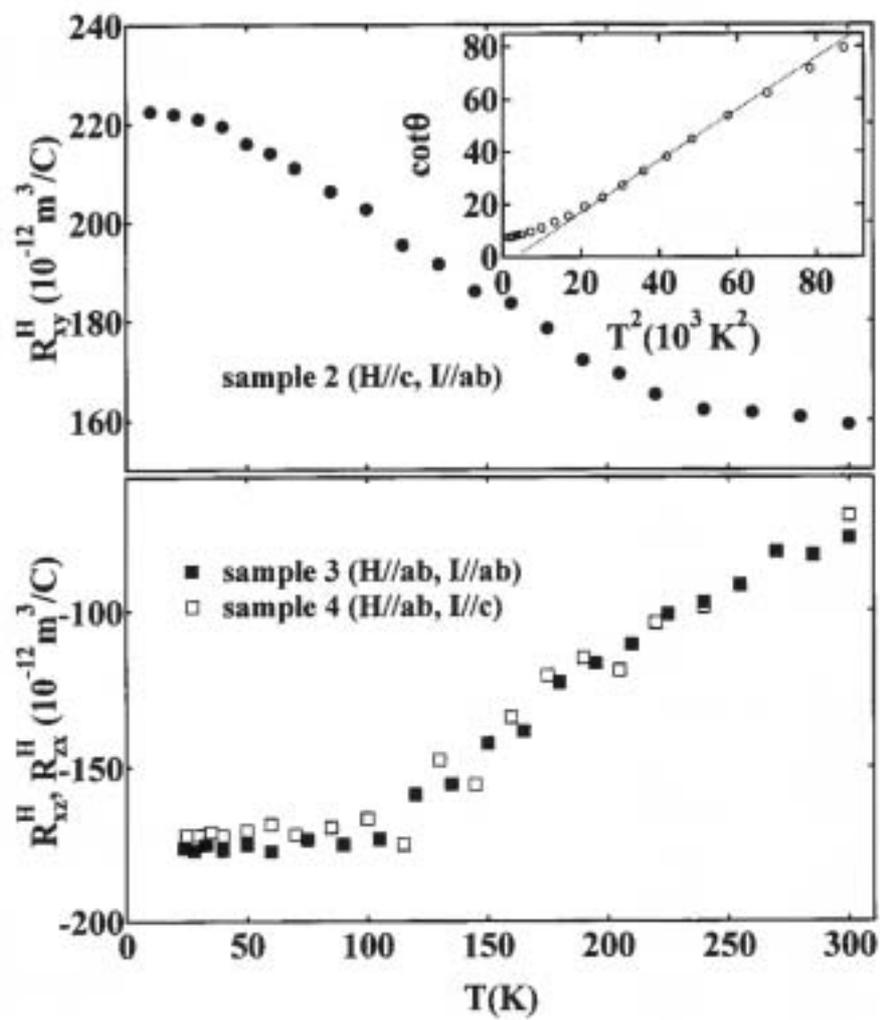

Fig. 9

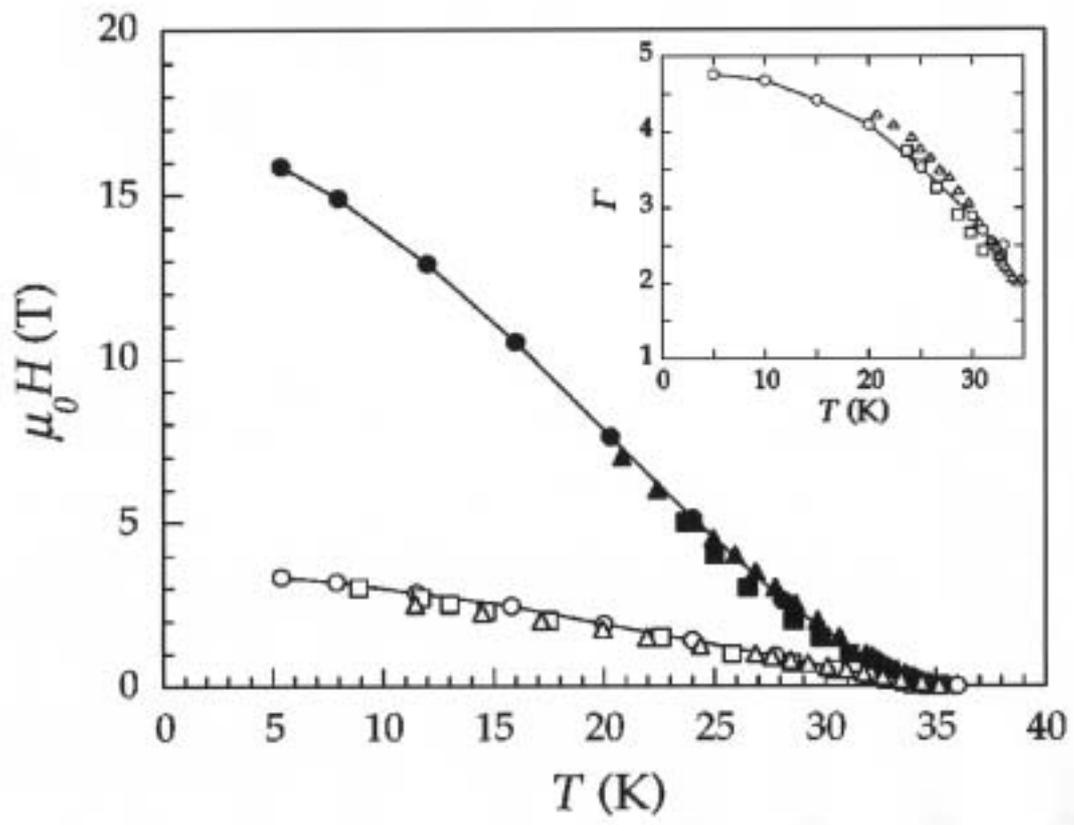

**Fig. 10**

**(a)**

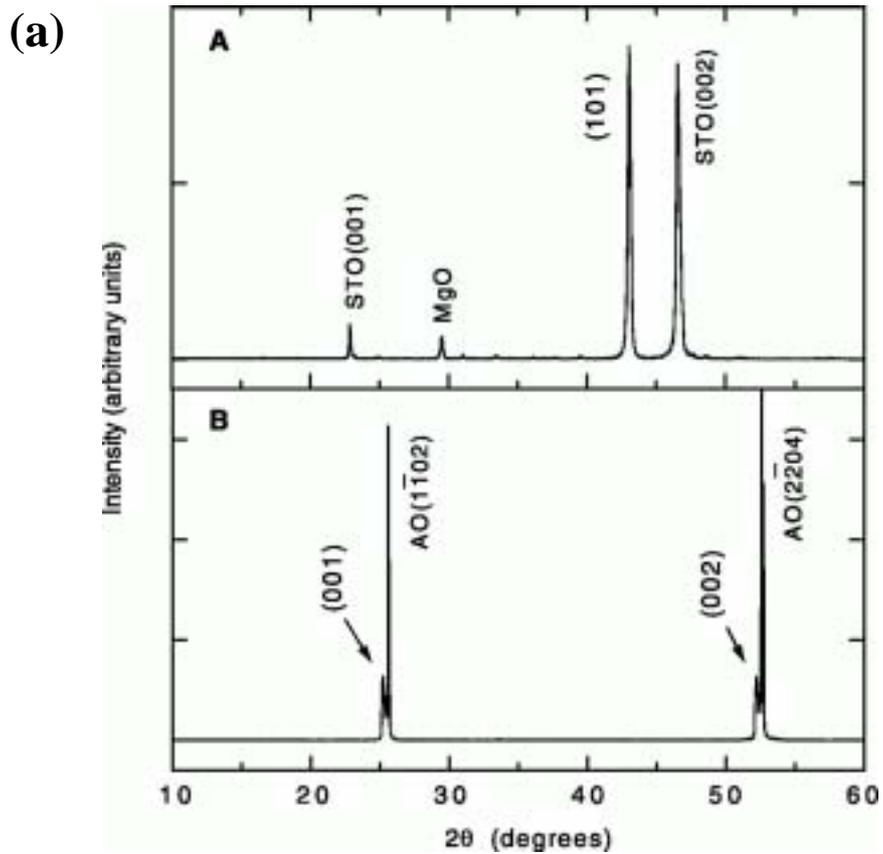

**(b)**

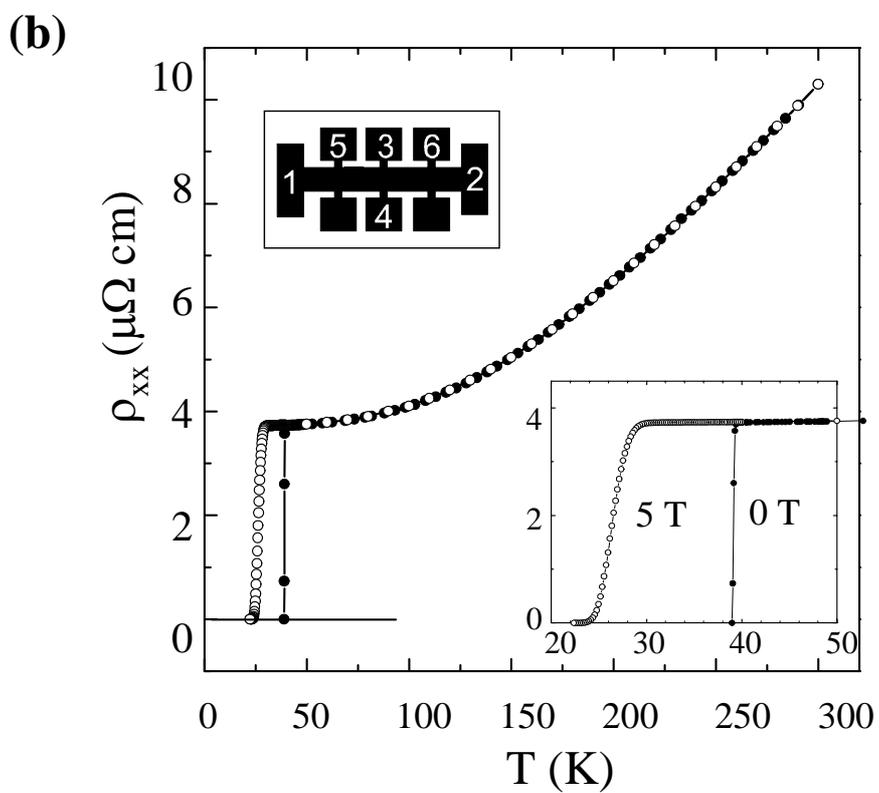

**Fig. 11**

**(a)**

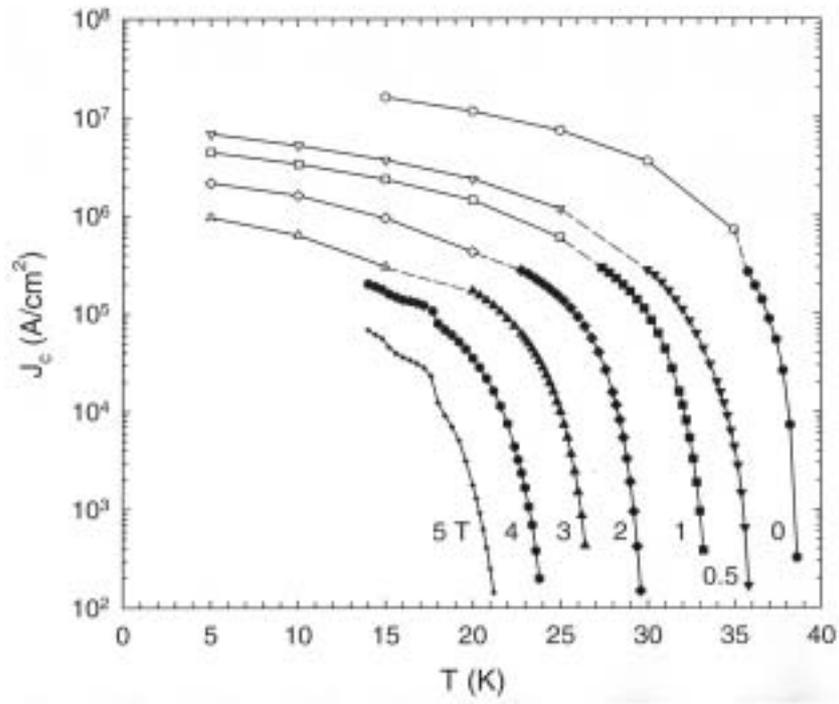

**(b)**

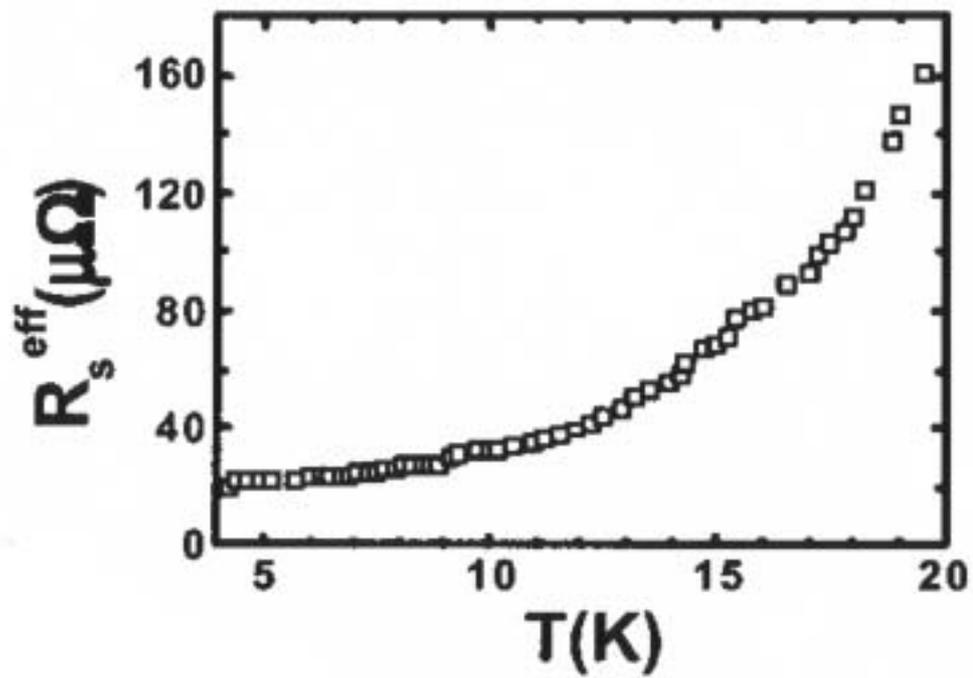

**Fig. 12**

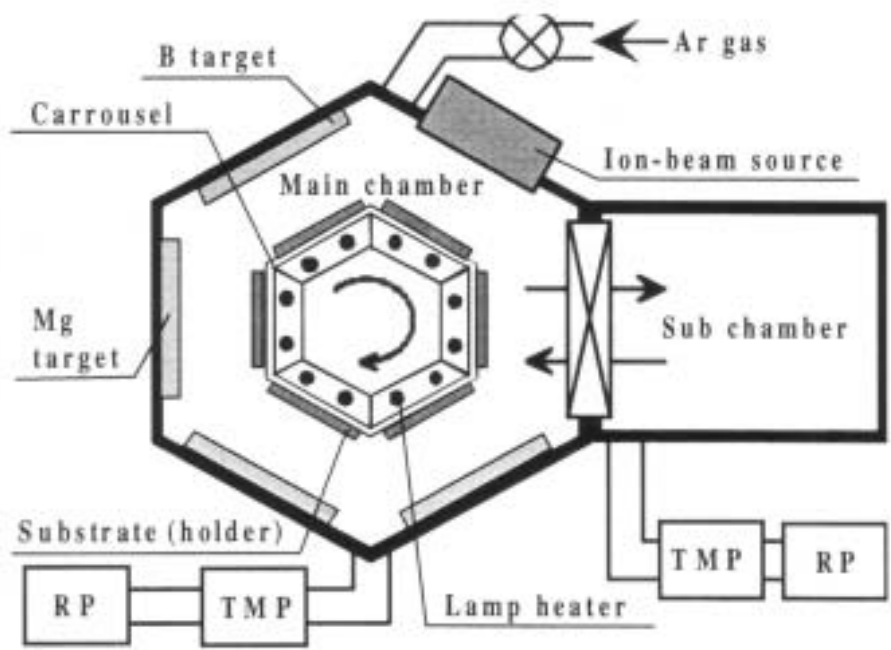

**Fig. 13**

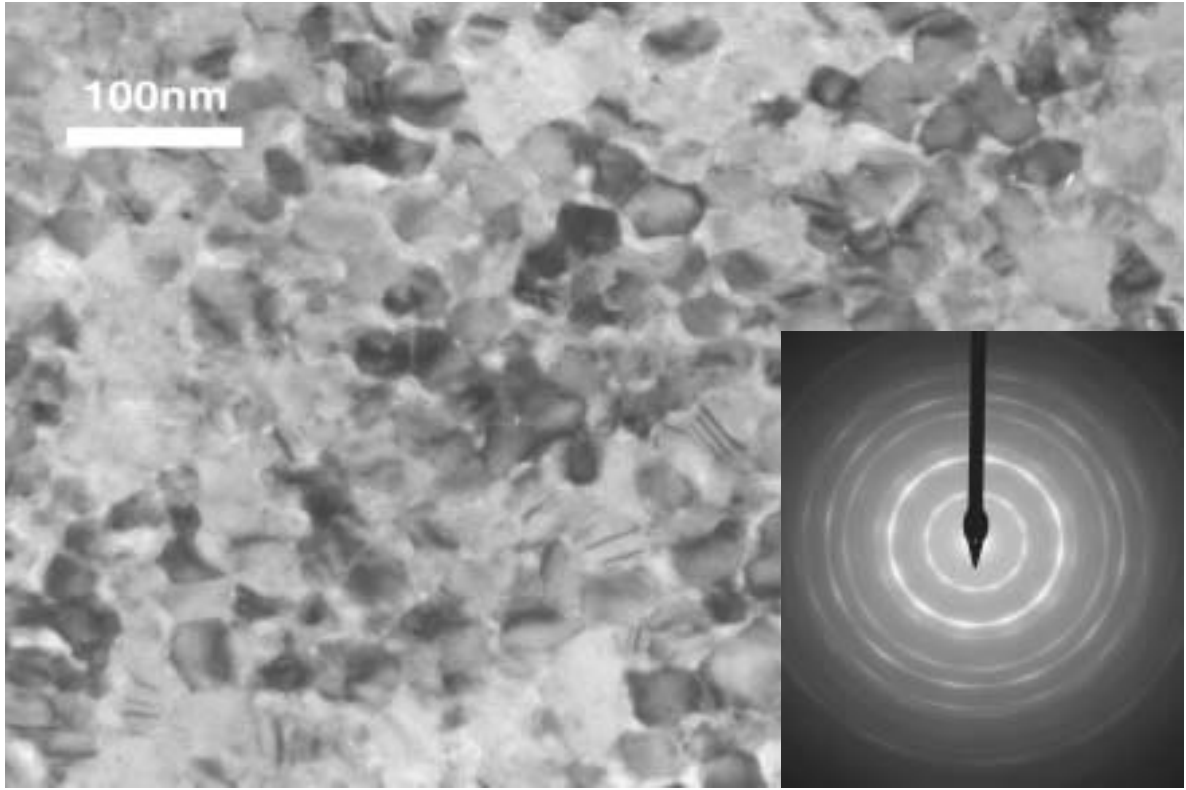

**Fig. 14**

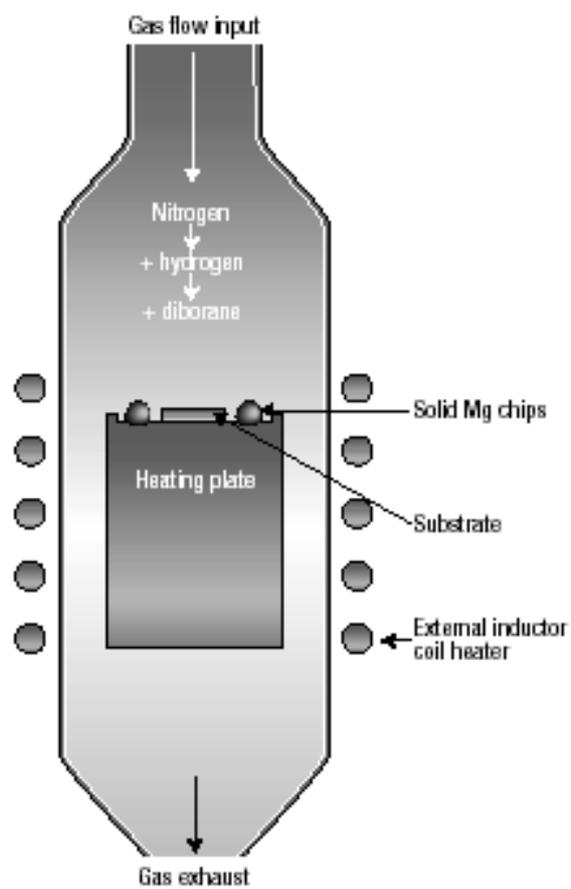

**Fig. 15**

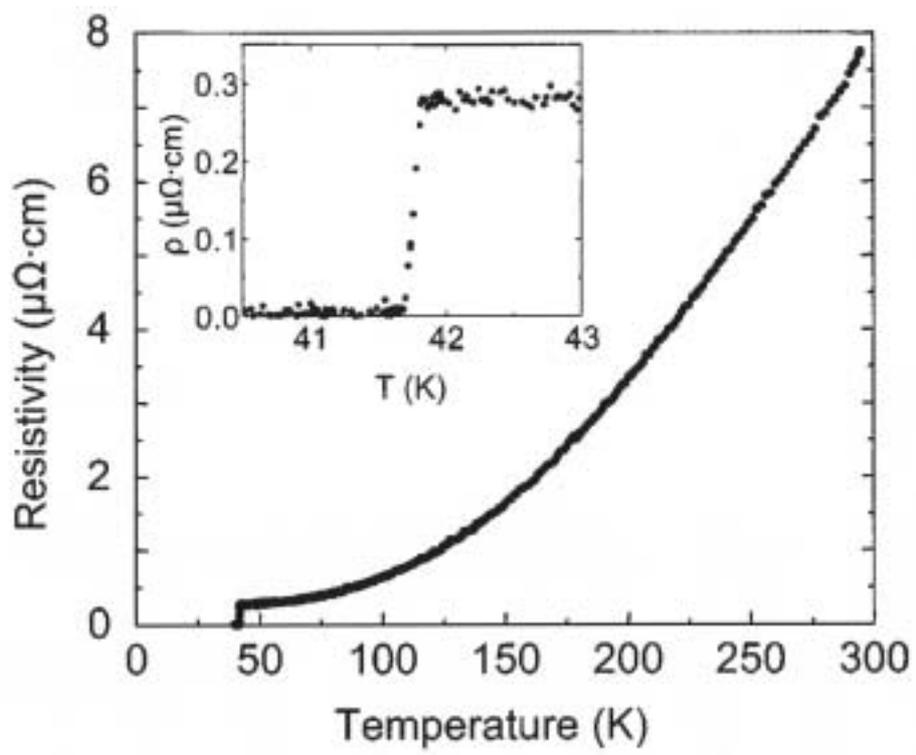

**Fig. 16**

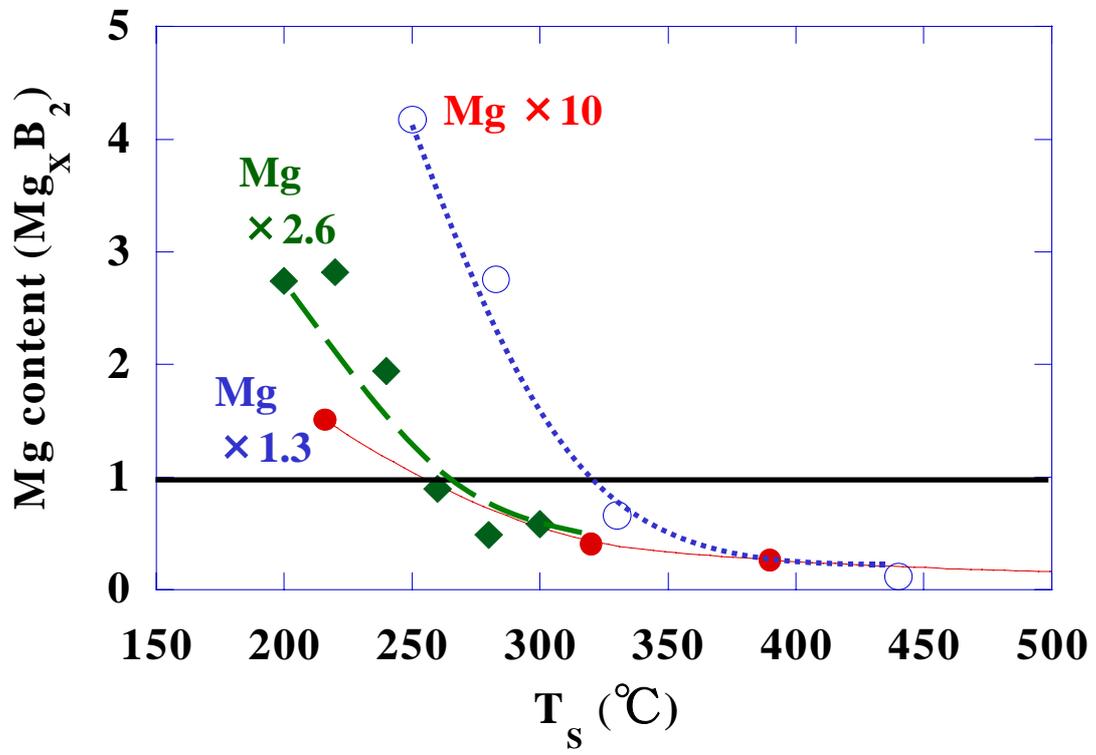

Fig. 17

(a)

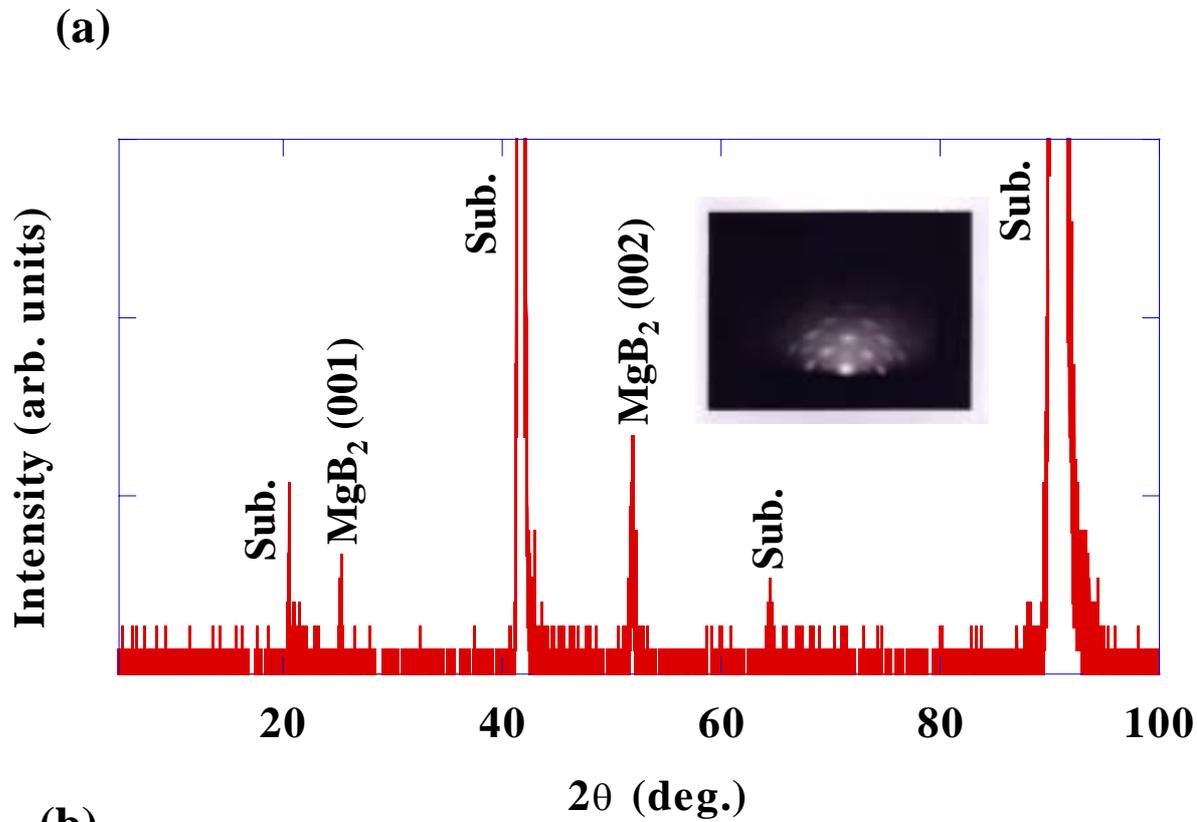

(b)

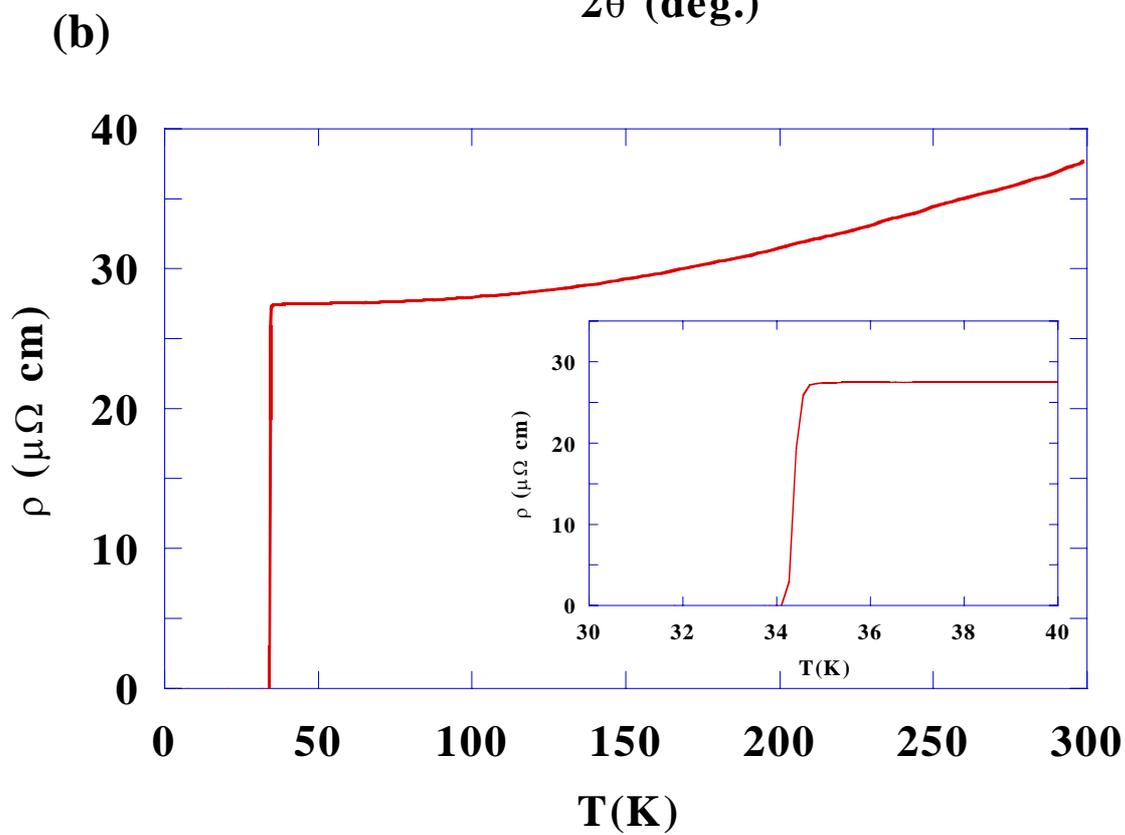

Fig. 18

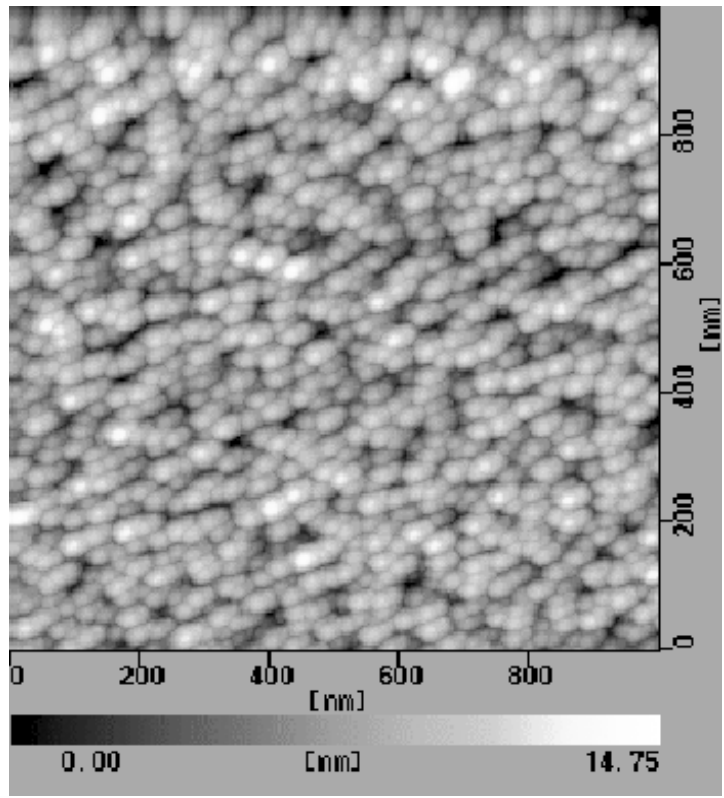

**Fig. 19**

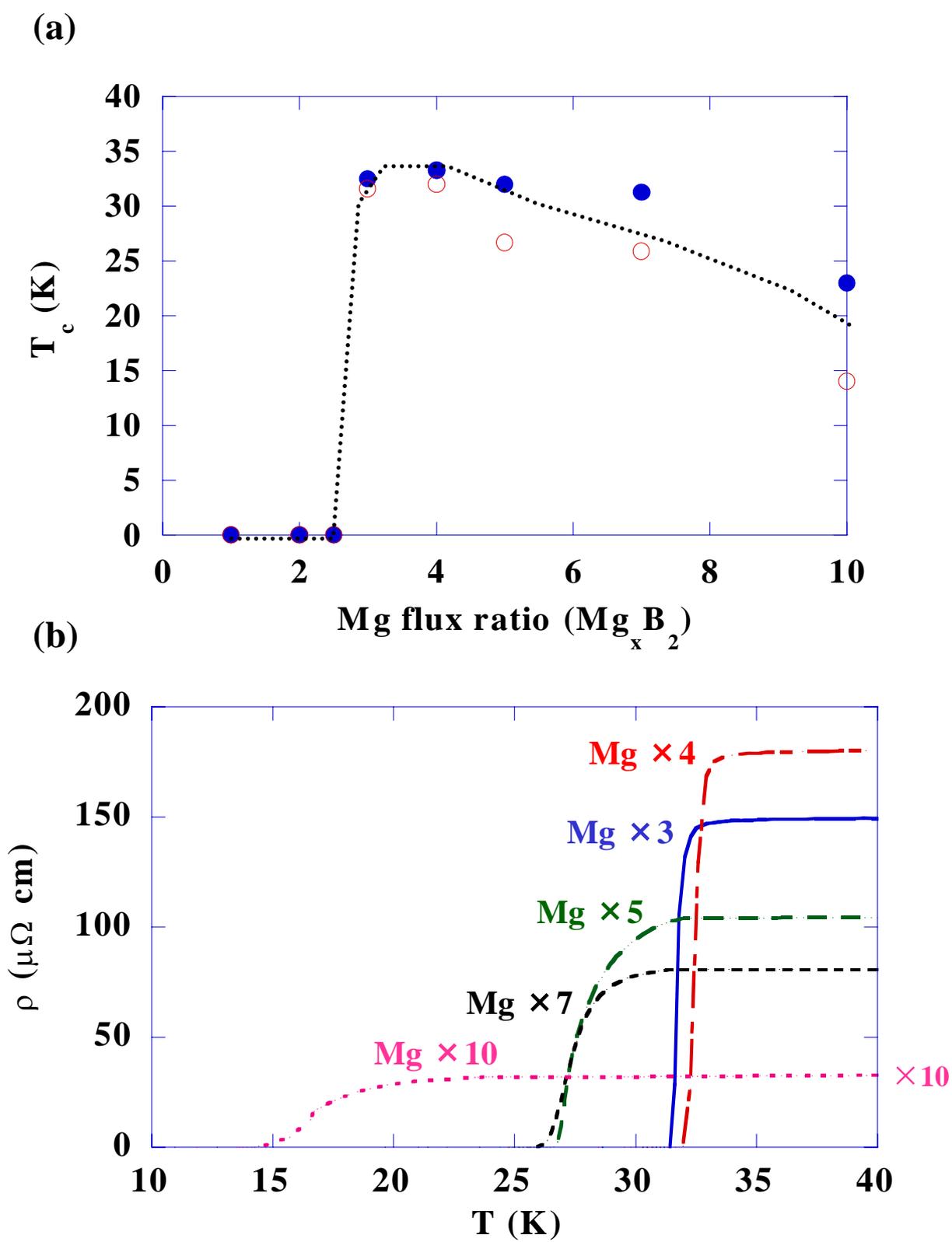

Fig. 20

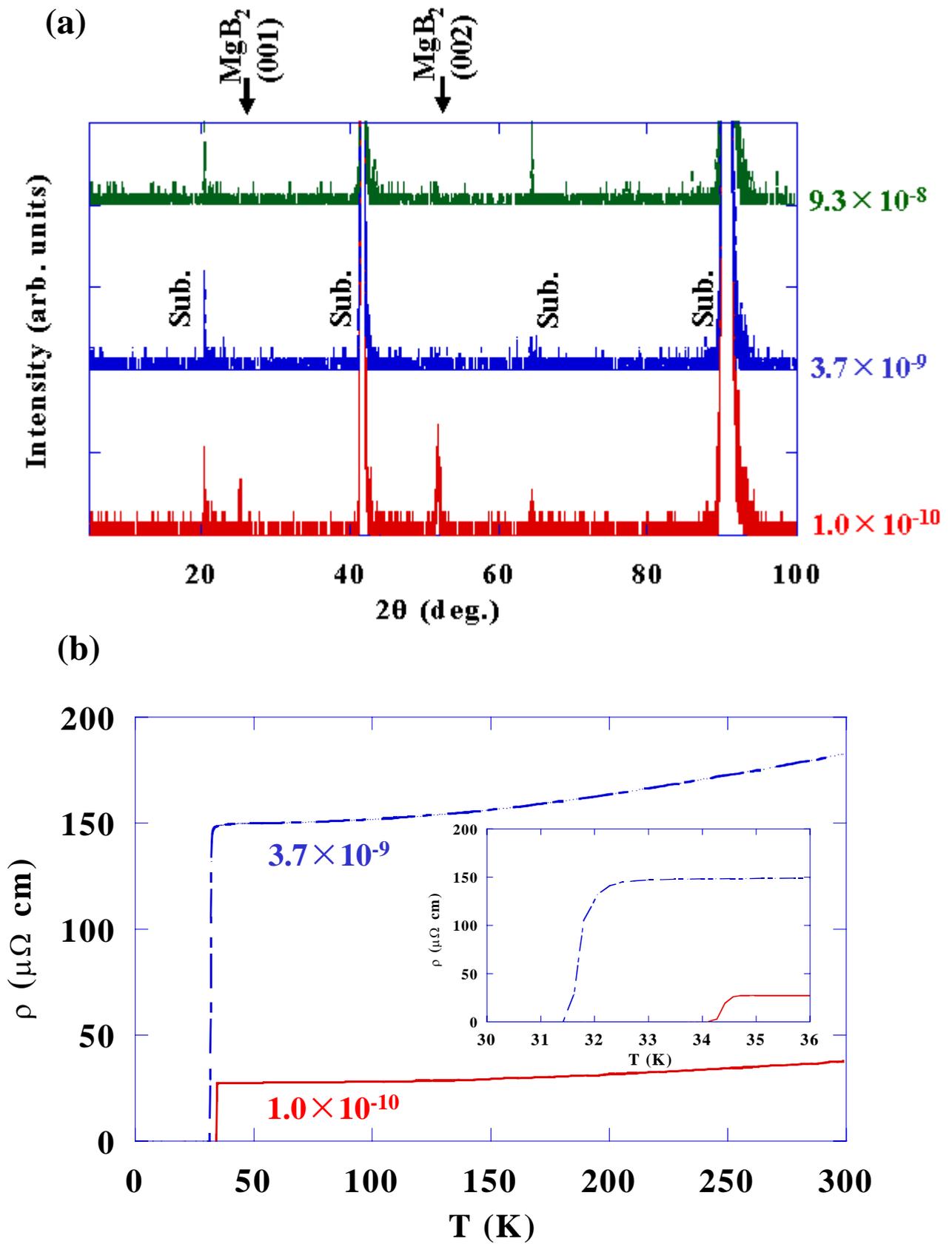

Fig. 21

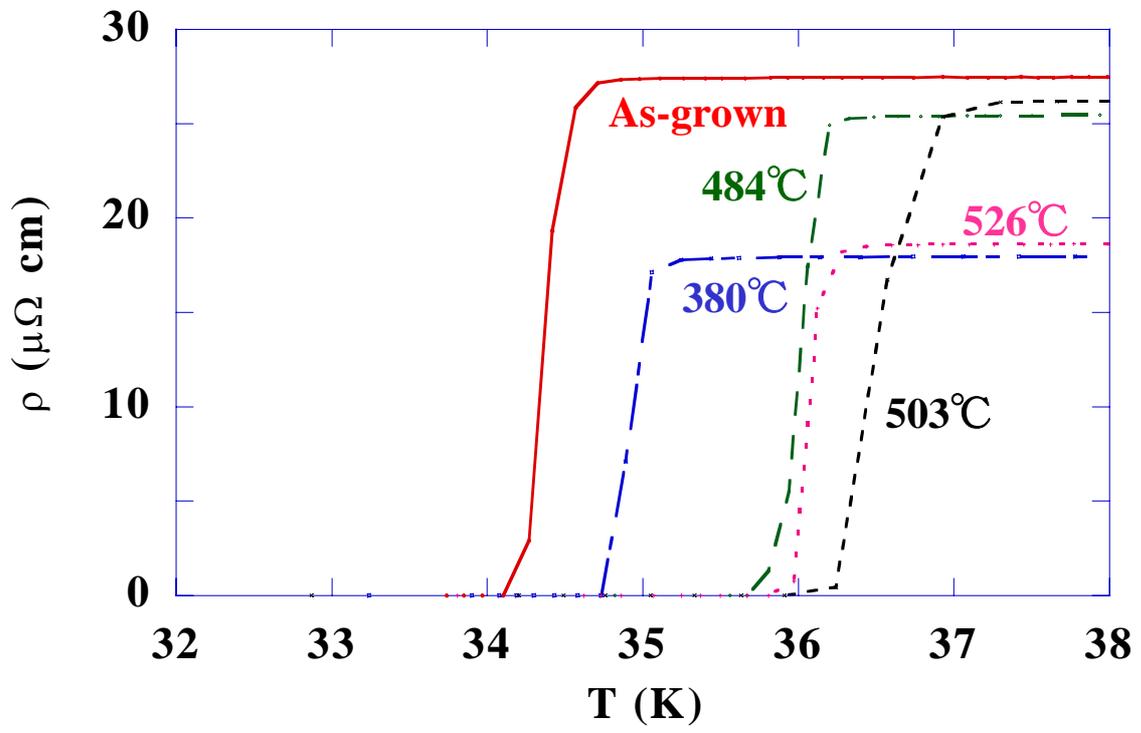

**Fig. 22**

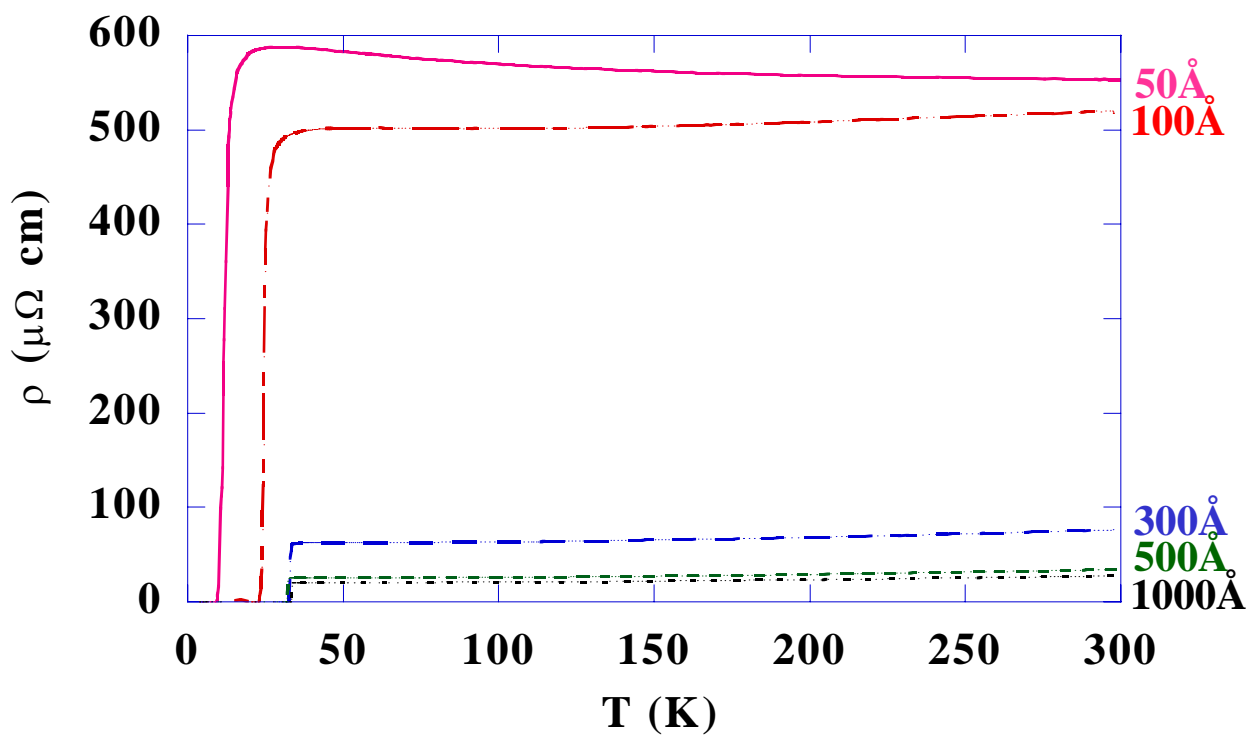

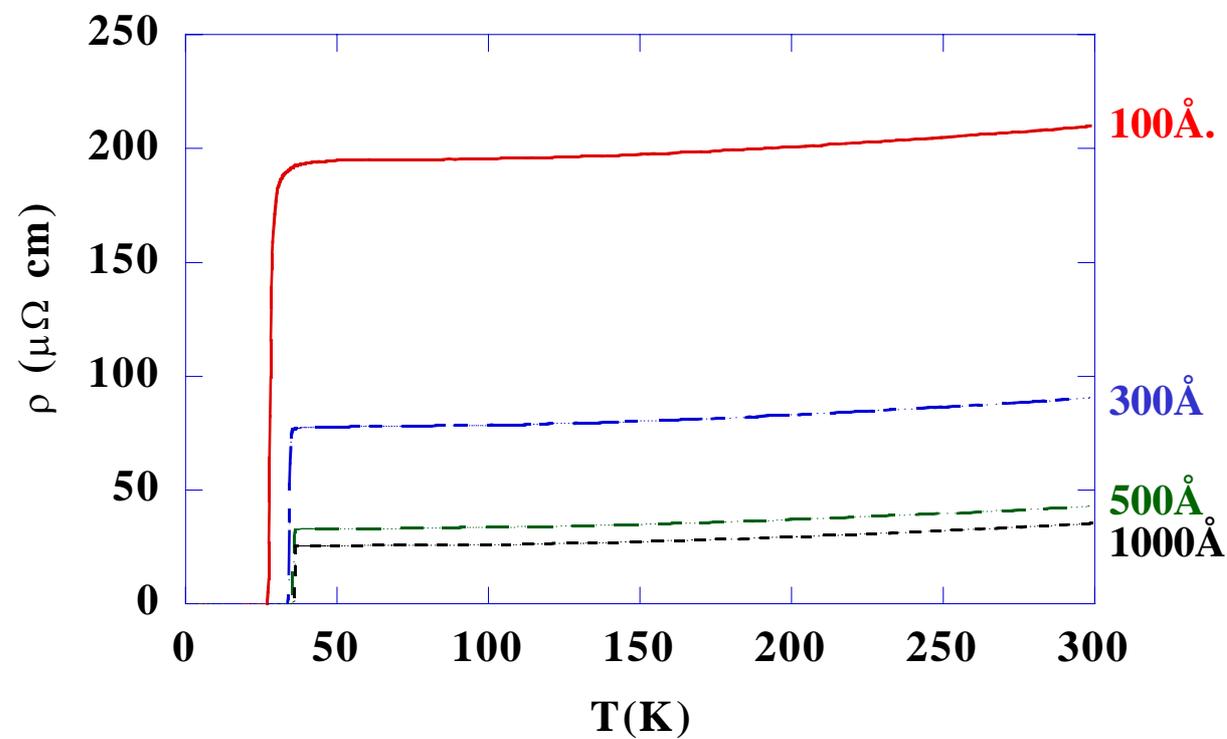

**Fig. 23**

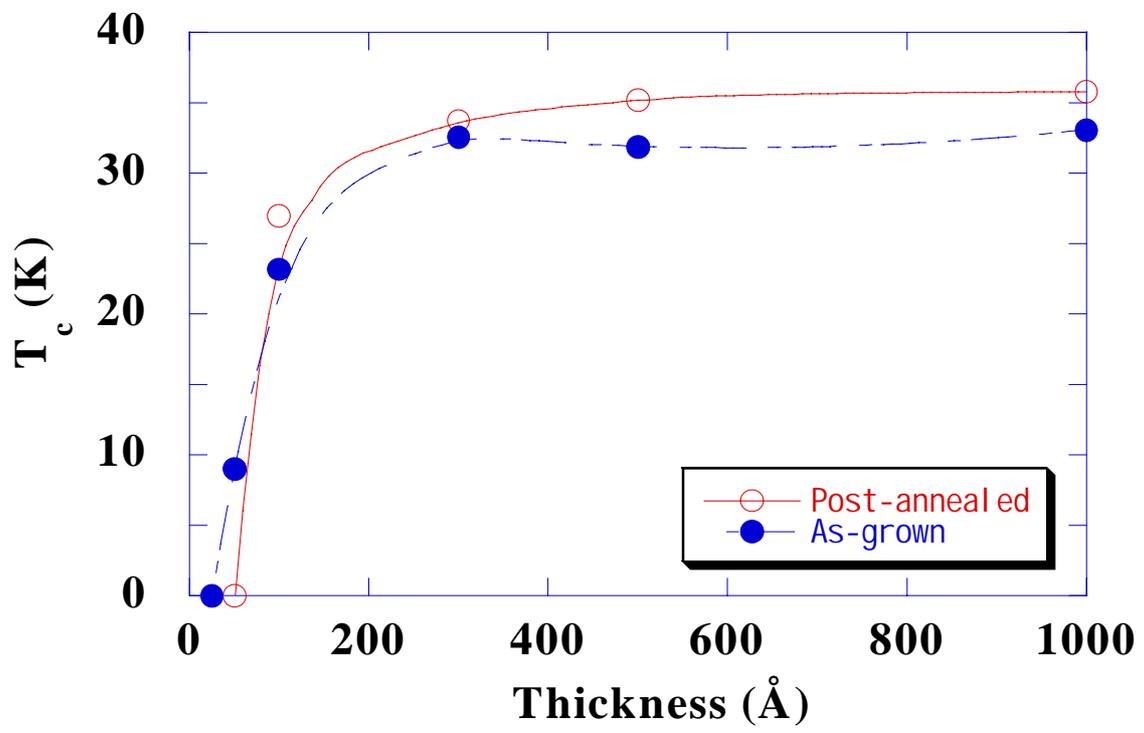

**Fig. 24**

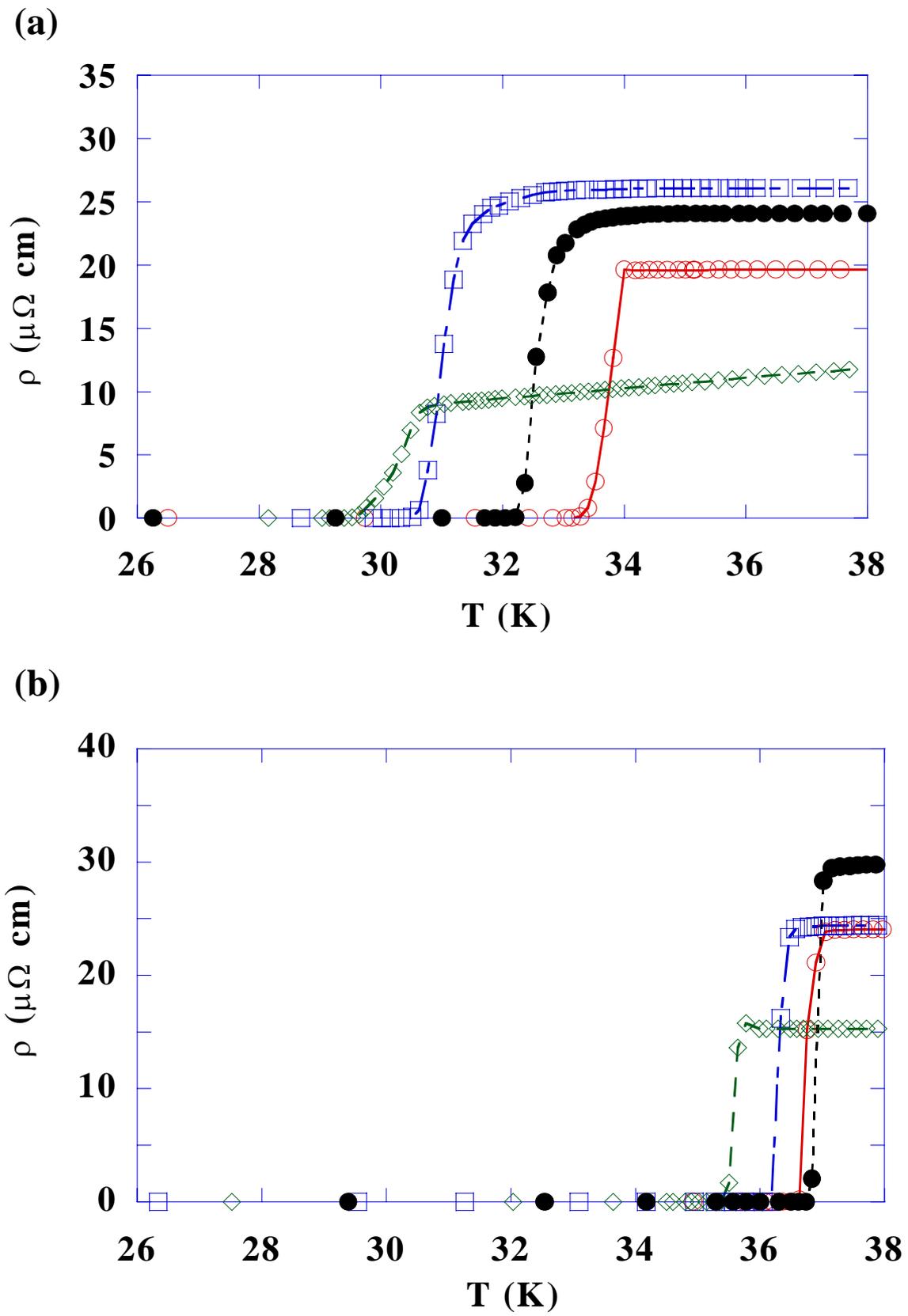

Fig. 25

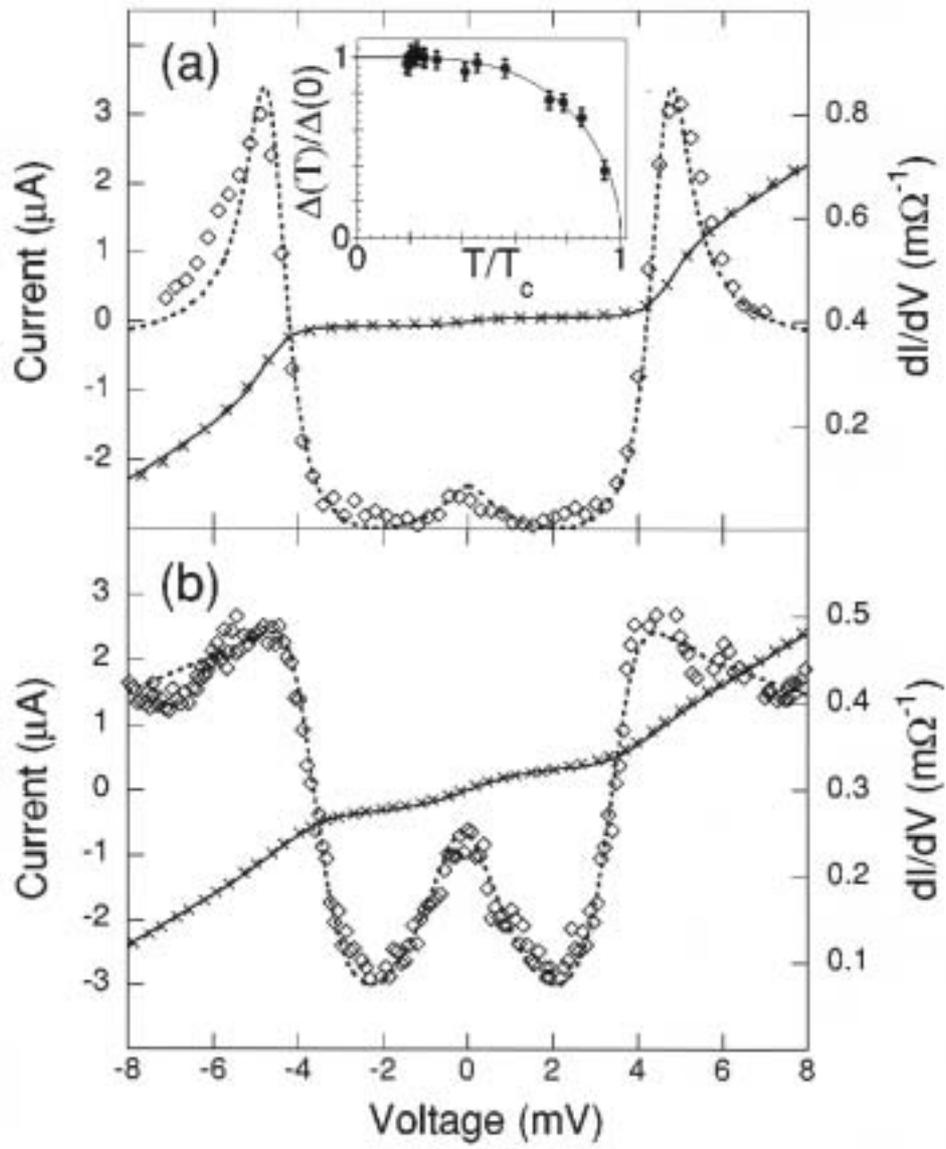

**Fig. 26**

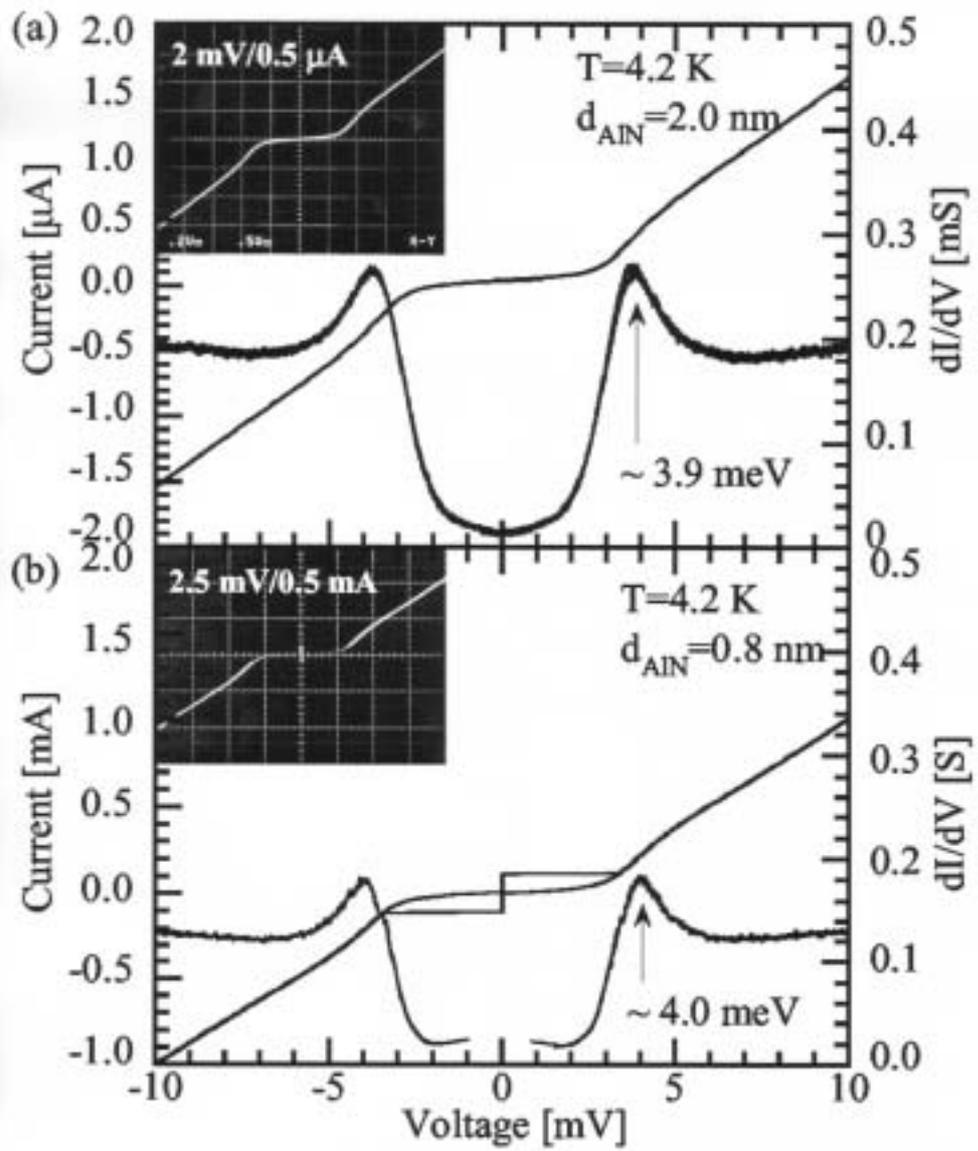
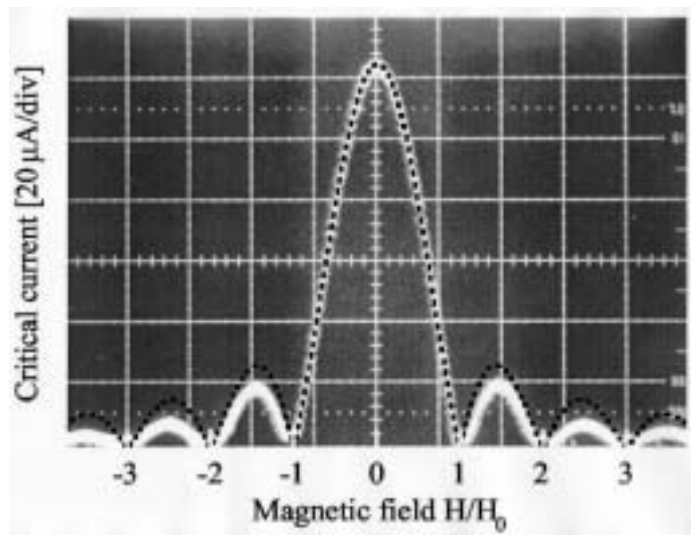

Fig. 27

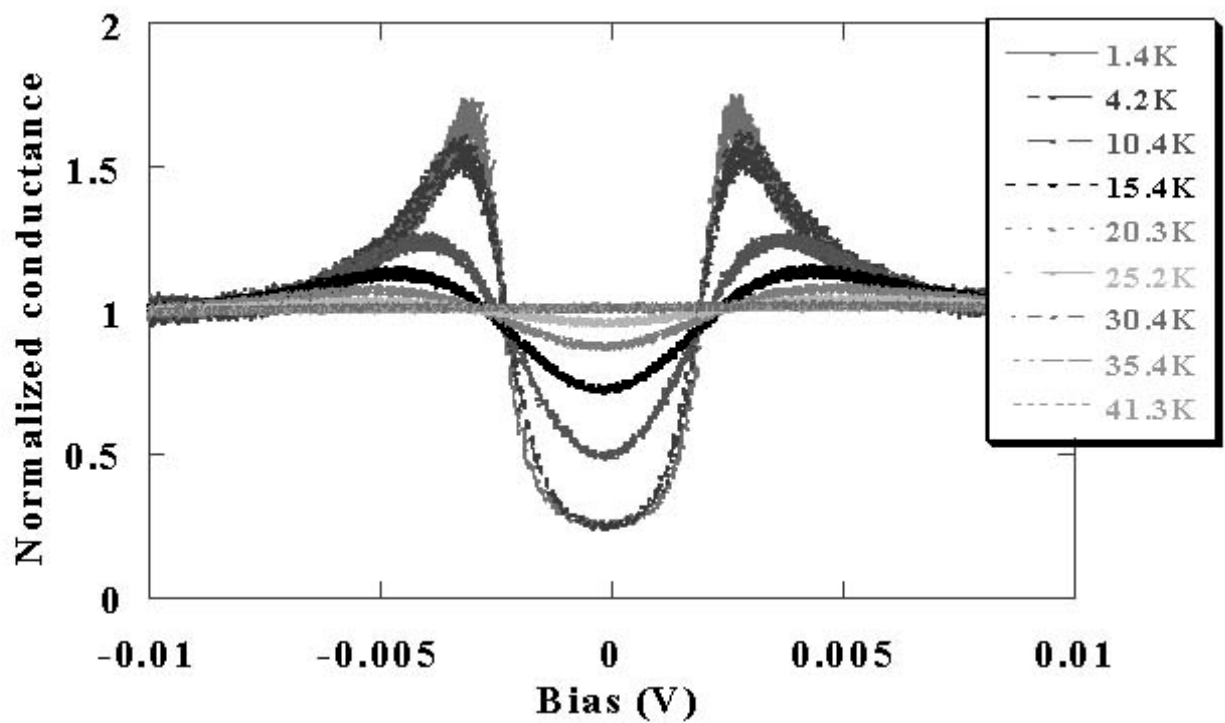

**Fig. 28**

# Table I

| Ref | Authors | Method | Precursors | Substrates | Thickness (mm) | Temp. (°C) | Time | Anneal tube | Tc(K) | RRR | Jc(A/cm$^2$) | XRD orientation |
|---|---|---|---|---|---|---|---|---|---|---|---|---|
| 39 | Kang | PLD | B-film | SrTiO$_3$-(100) | 0.4 | 900 | 10-30 min. | Ta | 39-37.6 | 2.5 | 6*10$^6$ (5K, 0T) | (101) |
|  |  |  |  | Al$_2$O$_3$--R |  |  |  |  |  |  |  | (001) |
| 40 | Eom | PLD | Mg+-B-film | SrTiO$_3$-(111) |  | 850 | 15 min. | Nb | 36-34 | ~1 | 3*10$^6$ (4.2K, 1T) | (001) |
|  |  |  |  |  |  | 750 | 30 min | Ta in Nb | 34-30 |  |  |  |
|  |  |  |  |  |  | 750 | 30 min | Quartz only | 34-29 |  |  |  |
| 41 | Paranthaman | E-beam | B-film | Al$_2$O$_3$--R | 0.5-0.6 | 890 | 10-20 min | Ta | 39-38.6 |  | 2*10$^6$ (20K, 0T) | (001) |
| 42 | Moon | E-beam | B-film | Al$_2$O$_3$--C MgO | 0.25-0.3 | 700--950 | 30 min | Ta, Ti | 39.2-38.9 | 1.5 | 1.1*10$^7$ (15K,0T) | Poly |
| 43 | Zhai | E-beam | B-film |  |  |  |  |  | 39-38.8 |  |  |  |
|  |  | PLD | Mg+-B-film | Al$_2$O$_3$--R |  | 900 | 1 hr | Ta | 28.6-25.2 |  |  |  |

# Table II

| Ref. | First Author | Method | Source | Substrates | Growth temp.(°C) | $T_c$ (K) | RRR | $J_c$(A/cm$^2$) | Others |
|---|---|---|---|---|---|---|---|---|---|
| 57 | Ueda | MBE | Mg and B metal | Al$_2$O$_3$, SrTiO$_3$, Si | 280 | 36-33.5 | ~1.5 | 4.0*10$^5$ (4.2K, 0T) | Co-deposition |
| 58,67 | Saito | sputtering | Mg, B two target | Al$_2$O$_3$ | 252 | ~28 | 2 | | Carrousel sputtering |
| 59 | Grassano | PLD | Mg+B pressed pellet | Al$_2$O$_3$, MgO | 400-450 | ~25 | ~1 | | Mg plasma |
| 60 | Jo | MBE | Mg and B metal | Al$_2$O$_3$ | 295-300 | 34.5-34 | 8 | | Co-deposition |
| 61 | Erven | MBE | Mg and B metal | Si | 300 | 34.9 | ~1.5 | | MgO seed layer |
| 62-65 | Zeng | HV-PVCVD | Mg metal and B$_2$H$_6$ | SiC, Al$_2$O$_3$ | 720-760 | 39.3-41.3 | 12-13 | 3.5*10$^7$ (4.2K, 0T) | Epitaxial films |

**Table III**

| Sub. | Crystal system | a (Å) | c (Å) | Surface | (lattice constant: Å) |
|---|---|---|---|---|---|
| SrTiO$_3$ | Cubic | 3.905 | | (100) | Square (3.905) |
| MgO | Cubic | 4.21 | | (100) | Square (4.21) |
| Si | Cubic | 5.431 | | (100) | Square (5.43) |
| Sapphire (Al$_2$O$_3$) | Hex. | 4.76 | 12.99 | (111) | Hex. (3.84) |
| | | | | C | Hex. (4.76) |
| | | | | R | Rectangle (4.76×15.39) |
| SiC | Hex. | 3.081 | 15.12 | (001) | Hex (3.081) |
| MgB$_2$ | Hex. | 3.086 | 3.522 | (001) | — |